\documentclass[aps,pre,twocolumn,showpacs,amsfonts,
eqsecnum,superscriptaddress]{revtex4}
\usepackage{graphicx}
\usepackage{epsfig}
\usepackage{bm}

\newcommand{\Onecol} {\begin{widetext} \onecolumngrid} 
\newcommand{\Twocol} {\end{widetext} \twocolumngrid}   
\newcommand{\Ref}[1]{(\ref{eq:#1})}
\newcommand{\REF}[1]{Eq.~(\ref{eq:#1})}
 
   \def\ie{{{i.e.}}}
 \renewcommand{\it}[1]{\textit{#1}}
\renewcommand{\S}[2]{\section{\label{s:#1}{#2}}}
\renewcommand{\SS}[2]{\subsection{\label{ss:#1}{#2}}}
\newcommand{\SSS}[2]{\subsubsection{\label{sss:#1}{#2}}}
\newcommand{\BE}[1]{\begin{equation}\label{eq:#1}}    
\newcommand{\EE}{\end{equation}}
\newcommand{\BEA}[1]{\begin{eqnarray} \label{eq:#1}} 
\newcommand{\EEA}{\end{eqnarray}} 
\newcommand{\BEa}{\begin{eqnarray*}} 
\newcommand{\EEa}{\end{eqnarray*}}   
\def\nn {\nonumber} \def\br {\\ \nonumber} 
\newcommand{\B}[1]{{\bm{#1}}}
\newcommand{\C}[1]{{\mathcal{#1}}}    
 \def\r{\B {r}} 
\def\d{\text{d}} 
\newcommand{\p}{\partial}           
\def\hf{\frac{1}{2}} 
\def\la{\langle} \def\ra{\rangle} 
\def\lla{\langle\!\langle} 
\def\rra{\rangle\!\rangle} 

\def\<{\left \langle} \def\>{\right\rangle}
\def\Re{{\C R}\mkern-3.1mu e}  

\newcommand{\tr}{(t,\B r)}

\renewcommand{\sb}[1]{_{\text {#1}}}  
\renewcommand{\sp}[1]{^{\text {#1}}}  

\def\Fbox#1{\vskip1ex\hbox to 8.5cm{\hfil\fboxsep0.3cm\fbox{%
  \parbox{8.0cm}{#1}}\hfil}\vskip1ex\noindent}  
\def\K41{\fboxrule0.2ex\fbox{\large\text{K41}}}

\def\Re{\, {\C R}\mkern-3.1mu e\,}

\begin{document}
\title{The clustering instability of inertial particles spatial
distribution in turbulent flows}
\author{Tov Elperin}
\email{elperin@menix.bgu.ac.il}
\homepage{http://www.bgu.ac.il/~elperin} \affiliation{ The
Pearlstone Center for Aeronautical
Engineering Studies, Department of Mechanical Engineering,\\
Ben-Gurion University of the Negev, Beer-Sheva 84105, P. O. Box
653, Israel}
\author{Nathan Kleeorin}
\email{nat@menix.bgu.ac.il}
\affiliation{ The Pearlstone Center for
Aeronautical Engineering
Studies, Department of Mechanical Engineering,\\
Ben-Gurion University of the Negev, Beer-Sheva 84105, P. O. Box
653, Israel}
\author{Victor S. L'vov}
\email{Victor.Lvov@Weizmann.ac.il}
\homepage{http://lvov.weizmann.ac.il}
\affiliation{ Department of
Chemical Physics, The Weizmann Institute of Science, Rehovot
76100, Israel}
\author{Igor Rogachevskii}
\email{gary@menix.bgu.ac.il} \homepage{http://www.bgu.ac.il/~gary}
\affiliation{ The Pearlstone Center for Aeronautical Engineering
Studies, Department of Mechanical Engineering,\\
Ben-Gurion University
of the Negev, Beer-Sheva 84105, P. O. Box 653, Israel}
\author{Dmitry Sokoloff}
\email{sokoloff@dds.srcc.msu.su}
\homepage{http://www.srcc.msu.su/lemg} \affiliation{ Department of
Physics,  Moscow State University, Moscow 117234, Russia}
\begin{abstract}
A theory of clustering of inertial particles advected by a
turbulent velocity field caused by an instability of their spatial
distribution is suggested. The reason for the clustering
instability is a combined effect of the particles inertia and a
finite correlation time of the velocity field. The crucial
parameter for the clustering instability is a size of the
particles. The critical size is estimated for a strong clustering
(with a finite fraction of particles in clusters) associated with
the growth of the mean absolute value of the particles number
density and for a weak clustering associated with the growth of
the second and higher moments. A new concept of compressibility of
the turbulent diffusion tensor caused by a finite correlation time
of an incompressible velocity field is introduced. In this model
of the velocity field, the field of Lagrangian trajectories is not
divergence-free. A mechanism of saturation of the clustering
instability associated with the particles collisions in the
clusters is suggested. Applications of the analyzed effects to the
dynamics of droplets in the turbulent atmosphere are discussed. An
estimated nonlinear level of the saturation of the droplets number
density in clouds exceeds by the orders of magnitude their mean
number density. The critical size of cloud droplets required for
clusters formation is more than $20 \mu$m.
\end{abstract}
\pacs{47.27.Qb, 05.40.-a}
\date{Version of \today}
\maketitle
\section{Introduction}
\label{intro}

Formation and evolution of aerosols and droplets inhomogeneities
(clusters) are of fundamental significance in many areas of
environmental sciences, physics of the atmosphere and meteorology
(e.g.,  smog and fog formation, rain formation), transport and
mixing in industrial turbulent flows (like spray drying,
pulverized-coal-fired furnaces, cyclone dust separation, abrasive
water-jet cutting) and in turbulent combustion (see, e.g.,
\cite{CST98,T77,C80,S86,EKPR97,SR98,EKRI98,VY00}). The reason is
that the direct, hydrodynamic, diffusional and thermal
interactions of particles in dense clusters strongly affect the
character of the involved phenomena. Thus, e.g., enhanced binary
collisions between cloud droplets in dense clusters can cause fast
broadening of droplet size spectrum and rain formation (see, e.g.,
\cite{VY00}). Another example is combustion of pulverized coal or
sprays whereby reaction rate of a single particle or a droplet
differs considerably from a reaction rate of a coal particle or a
droplet in a cluster (see, e.g., \cite{AR92,ARD94}).

Analysis of experimental data shows that spatial distributions of
droplets in clouds are strongly inhomogeneous (see, e.g.,
\cite{PB89,KM93,B99,KS01}). Small-scale inhomogeneities in
particles distribution were observed also in laboratory turbulent
flows \cite{SE91,EF94,FKE94,HL00}.

It is well-known that turbulence results in a relaxation of
inhomogeneities of concentration due to turbulent diffusion,
whereas the opposite process, e.g.,  a preferential concentration
(\it{clustering}) of droplets and particles in turbulent fluid
flow still remains poorly understood.

In this paper we suggest a theory of clustering of particles and
droplets in turbulent flows. The clusters of particles are formed due
to an instability of their spatial distribution suggested in
Ref.~\cite{EKRI96} and caused by a combined effect of a particle
inertia and a finite velocity correlation time. Particles inside
turbulent eddies are carried out to the boundary regions between them
by inertial forces.  This mechanism of the preferential concentration
acts in all scales of turbulence, increasing toward small scales. An
opposite process, a relaxation of clusters is caused by a
scale-dependent turbulent diffusion. The turbulent diffusion decreases
towards to smaller scales. Therefore, the clustering instability
dominates in the Kolmogorov inner scale $\eta$, which separates
inertial and viscous scales. Exponential growth of the number of
particles in the clusters is saturated by their collisions.

In our previous study \cite{EKRI96} we suggested and analyzed
qualitatively an idea that inertia of particles may lead to their
clustering.  Later this idea was questioned by our quantitative
analysis \cite{EKRS00,EKRS99} of the Kraichnan model of turbulent
advection of particles by the delta-correlated in time random velocity
field. It was proved that the clustering of inertial particles does
not occur in the Kraichnan model. The latter result may be considered
as counterexample.

The main quantitative result of the theory of clustering
instability of inertial particles, suggested in this paper, is
\it{the existence of  this instability} under some conditions that
we determined. We showed that the inertia of the particles is only
one of the \it{necessary conditions} for particles clustering in
turbulent flow. In the present study we found a \it{second
necessary condition} for the clustering instability: a finite
correlation time of the fluid velocity field which in the
suggested theory results in a nonzero divergence of the field of
Lagrangian trajectories. This time is equal zero in the above
mentioned Kraichnan model (see \cite{K68}) which was the reason
for the disappearance of the instability in this particular model.

In this study we used a model of the turbulent velocity field with
a finite correlation time which drastically changes the dynamics
of inertial particles. In the framework of this model of the
velocity field we rigorously derived the \it{sufficient
conditions} for the clustering instability. We demonstrated the
existence of the new phenomena of \it{strong} and \it{weak}
clustering of inertial particles in a turbulent flow. These two
types of the clustering instabilities have different physical
meaning and different physical consequences in various phenomena.
We computed also the instability thresholds which are different
for the strong and weak clustering instabilities.

\section{Qualitative analysis of strong and weak
clustering}
\label{clust}
\subsection{Basic equations in the continuous media
approximation}
\label{Basic}

In this study we used the equation for the number density $
n(t,{\B r}) $ of particles advected by a turbulent velocity field
$ {\B u}(t,{\B r}) $:
\begin{eqnarray}
\frac{\partial n\tr }{ \partial t} + \B \nabla \cdot[n\tr\,\B v\tr] =
D\, \Delta n\tr \,,
\label{eq:T1}
\end{eqnarray}
where $ D = k T / 6 \pi \nu \rho a $ is the coefficient of
molecular (Brownian) diffusion, $ \nu $ is the fluid kinematic
viscosity, $ \rho $ and $ T $ are the fluid density and
temperature, respectively, $ a $ is the radius of a particle and $
k $ is the Boltzmann constant. Due to inertia of particles their
velocity ${\B v}(t,{\B r})\ne {\B u}(t,{\B r}) ,$ e.g., the field
${\B v}(t,{\B r})$ is not divergence-free even for div~${\B u}=0$
(see \cite{EKRI96}). Equation (\ref{eq:T1}) implies conservation
of the total number of particles in a closed volume. Consider
\begin{eqnarray}
  \label{eq:def-Theta}
   \Theta(t,{\B r}) = n\tr - \bar n \,,
\end{eqnarray}
the deviation of $n\tr$ from the uniform mean number density of
particles $ \bar n $.  Equation for $ \Theta(t,{\B r})$ follows from
Eq.~(\ref{eq:T1}):
\begin{eqnarray}
  \label{W25}
   \frac{\partial\Theta\tr}{\partial t} &+& [\B v\tr \cdot \B \nabla]
  \Theta\tr \br
&=& - \Theta\tr \,\mbox{div}\,\B v\tr + D \Delta \Theta\tr \, .
\end{eqnarray}
Here we assumed that the mean particles velocity is zero. We also
neglected the term $ \propto \bar n\,\mbox{div}\,\B v $ describing
an effect of an external source of fluctuations. This term does
not affect the growth rate of the instability. In the present
study we investigate only the effect of self-excitation of the
clustering instability, and we do not consider an effect of the
source term on the dynamics of fluctuations. The source term $
\propto \bar n\,\mbox{div}\,\B v $ causes another type of
fluctuations of particle number density which are not related with
an instability and are localized in the maximum scale of turbulent
motions. A mechanism of these fluctuations is related with
perturbations of the mean number density of particles by a random
divergent velocity field. The magnitude of these fluctuations is
much lower than that of fluctuations which are caused by the
clustering instability.

In our qualitative analysis of the problem we use
Eq.~(\ref{W25}) written in a co-moving with a cluster reference
frame. Formally, this may be done using the
Belinicher-L'vov (BL) representation (for details, see
\cite{LP2,LP1}). Let $\B\xi\sb L(t_0,\B r|\,t)$ will be Lagrangian
trajectory of the reference point [in the particle velocity field
$\B v\tr$] located at $\B r$ at time $t_0$ and $\B \rho\sb L
({t_0, \B r}|t)$ be an increment of the trajectory:
\begin{eqnarray}
\label{eq:rL}
\B \rho\sb L (t_0,\r |t) &=& \int^{t}_{t_0} \B v [\tau,\B \xi \sb L
(t_0,\r | \tau)] \,\d \tau \,,
\br  
\B \xi \sb L (t_0,\r |t) &\equiv &\B r+\B \rho\sb L(t_0,\r |t)\ .
\end{eqnarray}
By definition $\B \rho\sb L (t_0,\r|\,t_0)=0$ and $\B \xi(t_0,\B
r|\,t_0)=\B r$.  Define as $\B r_0$ a position of a center of
a cluster at the ``initial'' time $t_0=0$ (for the brevity of notations
hereafter we skip the label $t_0$ ) and consider a ``co-moving''
reference frame with the position of the origin at $\B
\zeta_0(t)\equiv\B \zeta(\B r_0|\,t)$. Then BL velocity field
$\tilde {\B v}(\B r_0|t,\B r)$ and BL velocity difference $\B W(\B
r_0|t,\B r)$ are defined as
\begin{eqnarray}
    \label{eq:BL-not}
    \tilde{\B v}(\B r_0|t,\B r)& \equiv & \B v [t,\B r +\B \rho\sb L
    ({\B r}_0|t)] \,,\\  \label{eq:BL-dif}
 \B W(\B r_0|t,\B r)& \equiv & \tilde
    {\B v}(\B r_0|t,\B r)- \tilde{\B v}(\B r_0|t,\B r_0 )\ .
  \end{eqnarray}
Actually the BL representation is very similar to the Lagrangian
description of the velocity field. The difference between the two
representations is that in the Lagrangian representation one
follows the trajectory of every fluid particle $\B r+\B \rho\sb L
(\B r,t_0|t)$ (located at $\B r$ at time $t=t_0$), whereas in
the BL-representation there is a special initial point $\B r_0$
(in our case the initial position of the center of the cluster)
whose trajectory determines the new coordinate system (see
\cite{LP2,LP1}). With time the BL-field $ \tilde{\B v}(\B r_0|t,\B
r)$ becomes very different from the Lagrangian velocity field. It must
be
noted that the simultaneous correlators of both, the Lagrangian
and the BL-velocity fields, are identical to the simultaneous
correlators of the Eulerian velocity $\B v(\B r,t)$. The reason is
that for stationary statistics the simultaneous correlators do not
depend on $t$, and in particular one can assume $t=t_0$.

Similar to Eq.~(\ref{eq:BL-not}) let us introduce BL representation
for $\Theta\tr$:
\begin{eqnarray}
  \label{eq:BL-Theta}
  \tilde\Theta (\B r_0|t,\B r)\equiv \Theta [t,\B r +\B \rho\sb L
    ({\B r}_0|t)]\ .
\end{eqnarray}
In BL variables defined by Eqs. (\ref{eq:BL-not})-(\ref{eq:BL-Theta}),
Eq.~(\ref{W25}) reads:
\begin{eqnarray}\label{BL25}
&& \frac{\partial\tilde\Theta(\B r_0|t,\B r) }{\partial t} + [\B W
(\B r_0|t,\B r)\cdot \B \nabla ] \tilde \Theta (\B r_0|t,\B r)\\ &
= & - \tilde\Theta (\B r_0|t,\B r)\, \mbox{div}\,\B W (\B r_0|t,\B
r)+ D\, \Delta \,\tilde\Theta (\B r_0|t,\B r) \ . \nonumber
\end{eqnarray}
The difference between Eqs.~(\ref{W25}) and~(\ref{BL25}) is that
Eq.~(\ref{BL25}) involves only velocity
difference~(\ref{eq:BL-dif}) in which the velocity $\tilde{\B v}
(\B r_0|t,\B r_0)$ of the cluster center is subtracted.

\subsection{Rigid-cluster Approximation}
\label{uni-cluster}

Consider qualitatively a time evolution of different statistical moments

of the deviation $\Theta\tr$ defined by
  \begin{eqnarray}\label{def-M}
    \C M_q(t)\equiv \langle |\Theta\tr|^q \rangle_v\,,
  \end{eqnarray}
assuming that at the initial time, $t=0$, the spatial distribution
of particles is almost homogeneous, all moments $ \C M_q(0)$ are
small, where $ \langle \cdot \rangle_v $ denotes the  ensemble
averaging over random velocity field ${\B v}$.  In order to
eliminate the kinematic effect of sweeping of the cluster as a
whole we consider Eq.~(\ref{W25}) in the BL-representation,
Eq.~(\ref{BL25}). Since the simultaneous moments of any field
variables in the Eulerian and in the BL-representations coincide,
the moments $\C M_q(t)$ can be written as
 \begin{eqnarray}\label{def2-M}
    \C M_q(t)=\langle |\tilde\Theta(\B r_0|t,\B r)|^q \rangle_v\ .
  \end{eqnarray}
Our conjecture is that on a qualitative level we can consider the role
of
each term in the Eq.~(\ref{BL25}) separately, assuming some
reasonable, time independent, \it{frozen} shape $\theta(x)$ of a
distribution $ \tilde \Theta(\B r_0|t,\B r) $ inside a cluster:
\begin{eqnarray}
 \label{eq:shape}
 \tilde \Theta(\B r_0|t,\B r)
=A(t)\,  \theta\Big(\frac{|\B r-\B r_0|}{\ell\sb {cl}}\Big)\ .
 \end{eqnarray}
Here $A(t)$ is time-dependent amplitude of a cluster,
$\ell\sb{cl}$ is the characteristic width of the cluster. Shape
function $\theta(x)$ may be chosen with a maximum equal one at
$x=0$ and unit width.  Real shapes of various clusters in the
turbulent ensemble are determined by a competition of different
terms in the evolution equation~(\ref{BL25}).  However, we believe
that particular shapes affect only numerical factors in the
expression for the growth rate of clusters and do not effect their
functional dependence on the parameters of the problem which is
considered in this subsection.
\SSS{turb-dif}{Effect of turbulent diffusion}

The advective term in the LHS of Eq.~(\ref{BL25}) results in
turbulent diffusion inside the cluster. This effect may be
modelled by renormalization of the molecular diffusion coefficient
$D$ in the right hand side (RHS) of Eq.~(\ref{BL25}) by the
effective turbulent diffusion coefficient $D\sb T$ with a
usual estimate of $D\sb T$:
\begin{eqnarray}
  \label{eq:diff}
D\to D+ D\sb T\,,\quad D\sb T \sim \ell\sb{cl} v\sb{cl}/3 \ .
\end{eqnarray}
Hereafter $v\sb{cl}$ is the mean square velocity of particles
at the scale $ \ell\sb{cl} $. Instead of the full  Eq.~(\ref{BL25})
consider now a model equation
\begin{eqnarray}
\frac{\partial\tilde\Theta(\B r_0|t,\B r) }{\partial t}=D\sb T\,
\Delta \,\tilde\Theta (\B r_0|t,\B r) \,,
\label{mod5}
\end{eqnarray}
which accounts only for turbulent diffusion. Equations~(\ref{def2-M})
and (\ref{mod5}) yield:
\begin{eqnarray}
  \label{eq:mod2}
\hskip -0.5cm \partial \C M_q(t)/\partial t \simeq q \langle | \tilde
\Theta (\B r_0|t,\B r) |^{q-1} D\sb T \Delta |\tilde \Theta(\B
r_0|t,\B r) |\rangle_w .
\end{eqnarray}
Substituting distribution~(\ref{eq:shape}) we estimate
Laplacian in Eq.~(\ref{eq:mod2}) as $- 1/ \ell\sb{cl}^2$.
Equations~(\ref{mod5}) and (\ref{eq:mod2}) imply  that
\begin{eqnarray}
  \label{eq:mod3}
   \partial \C M_q(t)/\partial t = - q D\sb T\C M_q(t)/
   \ell\sb{cl}^2\ .
\end{eqnarray}
The solution of Eq.~(\ref{eq:mod3}) reads:
\begin{eqnarray}
\C M_q(t) &=& \C M_q(0) \exp[-\gamma\sb{dif}(q) \, t]\,, \,
\nonumber\\
\gamma_{\rm dif}(q) & \sim &  q D\sb T/ \ell\sb{cl}^2\ ,
\label{eq:mod4}
\end{eqnarray}
where $\gamma\sb{dif}(q)$ denotes a contribution to the damping
rate of $\C M_q(t)$ caused by turbulent diffusion.
\SSS{comp}{Effect of particles inertia}

In this subsection we show that the term $-\tilde \Theta\,
\mbox{div}\,\B W $ in the RHS of Eq.~(\ref{BL25}) can result in an
exponential growth of $\C M_q(t)\propto \exp [\gamma\sb{in}(q)\,t]
,$ i.e., in the instability. We denoted here the contribution to
the growth rate of $\C M_q(t)$, caused by the inertia of
particles, by $\gamma\sb{in}(q)$.  In order to evaluate
$\gamma\sb{in}(q)$ we neglect now in Eq.~(\ref{BL25}) both, the
convective term in the LHS of this equation ({\em i.e.} the the
turbulent velocity difference inside the cluster) and the
molecular diffusion term. The resulting equation reads:
\begin{eqnarray}
  \label{mod1}
  \frac{\partial\tilde\Theta(\B r_0|t,\B r)}{\partial t} =- \tilde
  \Theta(\B r_0|t,\B r)\,   \mbox{div}\,\B W (\B r_0|t,\B r)\ .
\end{eqnarray}
The main contribution to the BL-velocity difference $\B W (\B
r_0|t,\B r)$ in the RHS of this equation is due to the eddies with
size $\ell\sb{cl}$, the characteristic size of the cluster. Denote
by $v\sb{cl}$ the characteristic velocity of these eddies and by $
\tau_{v}\sim \ell\sb{cl}/v\sb{cl} $ the corresponding correlation
time. In our qualitative analysis we neglect the $\B r$ dependence
of $\mbox{div}\,\B W (\B r_0|t,\B r)$ inside the cluster and
consider the divergence in Eq.~(\ref{mod1}) as a random process
$b(t)$ with a correlation time $\tau_{v}$:
\begin{eqnarray}
  \label{eq:b}
  \mbox{div}\,\B W (\B r_0|t,\B r)\to b(t)\ .
\end{eqnarray}
Together with the decomposition~(\ref{eq:shape}) this yields the
following equation for the cluster amplitude $A(t)$:
\begin{eqnarray}
  \label{amp}
  \frac{\partial A(t)}{\partial t} =- A(t)\, b(t) \ .
\end{eqnarray}
The solution of Eq.~(\ref{amp}) reads:
\begin{eqnarray}
  \label{W7}
A(t)=A_{0}\,\exp[-I(t)]\,,\quad
  I(t) \equiv \int _0^t b(\tau)d \tau \ .
\end{eqnarray}
 Integral $I(t)$ in Eq.~(\ref{W7}) can be rewritten as a sum of
 integrals $I_n$ over small time intervals $ \tau_{v}$:
$$
I(t)= \sum_{n=1}^{t/\tau_v} I_n\,,\quad I_n(t)\equiv
\int\limits_{(n-1) \tau_v}^{ n \tau_v}b(\tau) \,d \tau\ .
$$
In our qualitative analysis integrals $I_n$ may be considered as
independent random variables.  Using the central limit theorem we
estimate the total integral
$$
I(t) \sim \sqrt{\langle I_{n}^{2} \rangle_v} \sqrt{N} \zeta\,,\quad
\langle I_{n}^{2} \rangle_v = \langle b^{2} \rangle_v \,
\tau_{v}^{2}\,,
$$
where $ \, \langle\dots \rangle_v$ denotes
averaging over turbulent velocity ensemble, $ \zeta $ is a Gaussian
random variable with zero mean and unit variance, $ N = t /
\tau_{v} .$ Now we calculate
$$
\C M_q(t) = \int \Theta^{q} P(\zeta) \,
d\zeta\,,\quad P(\zeta) = (1 / \sqrt{2 \pi}) \exp(- \zeta^{2} / 2)\ .
$$
Therefore, $\C M_q(t) = J_{1} \exp (q^{2} S^{2} N / 2) ,$ where
$$
J_{1} = (1 / \sqrt{2 \pi}) \int \exp[- (\zeta - q S \sqrt{N})^{2} / 2]
\,d\zeta \sim 1\ .
$$
Since the main contribution to the integral $ J_{1}$  arises from
$ \zeta \sim q S \sqrt{N} $, the parameter $ q $ cannot be large.
In this approximation the $q-$moments
$$
\C M_q(t)=\C M_q(0) \exp
  [\gamma\sb{in}(q) t]
$$
with  $\gamma\sb{in}(q)$ being the growth rate of the $q$-th
moment due to particles inertia which is given by
\begin{eqnarray}
  \label{eq:inst}
  \gamma\sb{in}(q)\sim \frac{1}{2} \langle \tau_{v} [\mbox{div} \,
  \B W(\B r_0|t,\B r)] ^{2} \rangle_v q^{2}\ .
\end{eqnarray}

\SSS{picture}{Qualitative picture of the clustering instability}

In previous subsections we evaluated the contributions to the
growth rate of $\C M_q(t)$ due to the turbulent diffusion
$\gamma\sb{dif}(q)$, Eq.~(\ref{eq:mod4}), and due to the particles
inertia $ \gamma\sb{in}(q)$, Eq.~(\ref{eq:inst}). The total growth
rate may be evaluated as a sum of these contributions:
\begin{eqnarray}
\label{gamma-q}
  \C M_q(t)&=& \C M_q(0)\exp (\gamma_q t)\,,\br
 \gamma_q &\simeq &
  \gamma\sb{dif}(q) +\gamma\sb{in}(q)\,, \br \gamma_q&\sim& - q
  D\sb T/\ell\sb{cl} + \frac{1}{2} \langle \tau_{v}
  [\text{div} \, \B W(\B r_0|t,\B r)] ^{2} \rangle_v q^{2}\ .
\end{eqnarray}
 Clearly, the instability is caused by a nonzero value of $
\langle\tau_{v} (\text{div} \, {\B W})^{2} \rangle ,$ {\em i.e.,} by a
{\em compressibility of the particle velocity field} $\B v(t,\B r)$.

Compressibility of fluid velocity itself $\B u(t,\B r)$ (including
atmospheric turbulence) is often negligible, i.e., $\text{div}\,\B
u \approx 0$. However, due to the effect of particles inertia
their velocity $\B v(t,\B r)$ does not coincide with $\B u(t,\B
r)$ (see, e.g., \cite{M87,MC86,WM93,MCW96}), and a degree of
compressibility,  $\sigma_{v}$,  of the field $\B v(t,\B r)$, may
be of the order of unity\cite{EKRI96,EKR96,EKRS00}.

Parameter $\sigma_{v}$ is defined as
\begin{eqnarray}
  \label{eq:def-s}
  \sigma_{v} \equiv \langle [\text{div} \, {\B v}]^2 \rangle / \langle
  |\B \nabla \times {\B v}|^{2} \rangle\ .
\end{eqnarray}
Note that parameter $\sigma_{v}$ is independent of the scale of
the turbulent velocity field. It characterizes a compressible part
of the velocity field as a whole. The main contribution to this
parameter comes from the scales which are of the order of the
Kolmogorov scale $\eta$.

For inertial particles $\text{div}\,\B v \sim \tau\sb{p} \Delta P
/ \rho $, where $ \tau\sb{p}$ is the particle response time,
\begin{eqnarray}
  \label{eq:tau-p}
  \tau\sb{p}= m\sb{p} /6\pi \rho \nu a =2 \rho\sb{p} a^2 /9
\rho \nu\,,
\end{eqnarray}
where $m\sb{p}$ and  $\rho\sb{p}$ are the mass and material
density of particles, respectively. The fluid flow parameters are:
pressure $P$, Reynolds number $\Re= L u\sb{T} / \nu$, the
dissipative scale of turbulence $ \eta = L {\Re}^{-3/4}$, the
maximum scale of turbulent motions $L$ and the turbulent velocity
$ u\sb{T}$ in the scale $L$. Now we can estimate $\sigma_{v}$ as
\begin{eqnarray}
\sigma_v\simeq (\rho\sb{p}/\rho)^2(a/\eta)^4\equiv (a/a_*)^4 \;
\label{W22}
\end{eqnarray}
(see \cite{EKRS00}), where $a_*$ is a characteristic radius of
particles. For $a>a_*$ it is plausible to correct this estimate as
follows:
\begin{eqnarray}
 \sigma_v\sim \frac{a^4}{a^4+a^4_*}\ .
\label{WW22}
\end{eqnarray}
For water droplets in the atmosphere $\rho\sb{p}/\rho\simeq 10^3$ and
$a_*\simeq \eta /30$. For the typical value of $\eta\simeq 1$mm it
yields $a_*\simeq 30\mu$m. On windy days when $\eta$ decreases, the
value of $a_*$ correspondingly becomes smaller.

Then we estimate $ \langle \tau_v [\text{div} \,\B v]^{2} \rangle$
as $ 2 \sigma_v/\tau_v$, because $ \langle [\text{rot} \,\B v]^{2}
\rangle \sim 2 \tau_v^{-2} $. Assuming that the cluster size $
\ell\sb{cl}$ is of the order of the inner scale of turbulence, $
\eta $, we have to identify $\tau_v$ with a turnover time of
eddies in the inner scale $\eta$, $\tau_v \to \tau_\eta \equiv
\eta/v_\eta=(L/ u\sb{T}) {\Re}^{-1/2}$. Thus, the growth rate $
\gamma_q$ of the $q$-th moment in Eq.~(\ref{gamma-q}) may be
evaluated as
\begin{eqnarray}
\gamma_q\simeq \gamma \sb{cl} \, q (q - q\sb{cr}) \,,\quad \gamma
\sb{cl} \sim {\sigma_v \over \tau _\eta} \,, \quad q\sb{cr} \sim
{1 \over 3 \sigma_v} \ . \label{W27}
\end{eqnarray}
Clearly, the moments with $q > q\sb{cr}$ are unstable. Equations
(\ref{W22}) and (\ref{W27}) imply that it happens when $
a>a_{q,\rm cr}$ where $a_{q,\rm cr}= a_{1,\rm cr}/ q^{1/4}$ is the
value of $a$ at which $q\sb{cr}=q$. The largest value of $a_{q,\rm
cr}$ corresponds to the instability of the first moment, $ \langle
|\Theta|\rangle$: $a_{1,\rm cr} \sim 0.8 \,a_*$, $a_{2,\rm cr}
\approx 0.84 \,a_{1,\rm cr}$, $a_{3,\rm cr} \approx 0.76
\,a_{1,\rm cr}$, $a_{4\rm cr} \approx 0.71 \, a_{1,\rm cr}$, {\em
etc.}

Note that if $ \langle |\Theta| \rangle $ grows in time then
almost all particles can be accumulated inside the clusters (if we
neglect a nonlinear saturation of such growth). We define this
case as a {\em strong clustering}. On the other hand, if $
q\sb{cr} > 1 $ the first moment $ \langle |\Theta| \rangle $ does
not grow, and the clusters contain a small fraction of the total
number of particles. This does not mean that the instability of
higher moments is not important. Thus, e.g., the rate of binary
particles collisions is proportional to the square of their number
density $\langle n ^2 \rangle=(\bar n)^2+ \langle |\Theta|^2
\rangle$. Therefore, the growth of the 2nd moment, $ \langle
|\Theta|^{2} \rangle$, (which we define as a {\em weak
clustering}) results in that binary collisions occur mainly
between particles inside the cluster. The latter can be important
in coagulation of droplets in atmospheric clouds whereby the
collisions between droplets play a crucial role in a rain
formation.  The growth of the $q$-th moment, $ \langle |\Theta|^q
\rangle$, results in that $q$-particles collisions occur mainly
between particles inside the cluster. The growth of the negative
moments of particles number density (possibly associated with
formation of voids and cellular structures) was discussed in
\cite{BF01} (see also \cite{SZ89,KS97}).

In the above qualitative analysis whereby we considered only one-point
correlation functions of the number density of particles, we missed an
important effect of an effective drift velocity which decreases a
growth rate of the clustering instability. For the one-point
correlation functions of the number density of particles the effective
drift velocity is zero for homogeneous and isotropic
turbulence. However, in the equations for two-point and multi-point
correlation functions of the number density of particles the effective
drift velocity is not zero and as we will see in the next section it
increases a threshold for the clustering instability.
\section{The clustering instability of the 2nd moment}
\label{moments}
\subsection{Basic equations}
\label{bas}

In the previous section we estimated the growth rates of all
moments $\langle |\Theta|^q \rangle$.  Here we present the results
of a rigorous analysis of the evolution of the two-point 2nd
moment
\begin{eqnarray}
  \label{eq:def-F}
  \Phi(t, \B R) \equiv \langle \Theta(t, \B r) \Theta(t, \B r+\B R )
  \rangle\ .
\end{eqnarray}
In this analysis we used stochastic calculus [e.g.,  Wiener path
integral representation of the solution of the Cauchy problem for
Eq.~(\ref{eq:T1}), Feynman-Kac formula and Cameron-Martin-Girsanov
theorem]. The comprehensive description of this approach can be found
in \cite{EKRS99,ZRS90,ZMR88,EKR00}.

We showed that a finite correlation time of a turbulent velocity
plays a crucial role for the clustering instability. Notably, an
equation for the second moment $\Phi(t, \B R)$ of the number
density of inertial particles comprises spatial derivatives of
high orders due to a non-local nature of turbulent transport of
inertial particles in a random velocity field with a finite
correlation time (see Appendix A and \cite{EKRS00}). However, we
found that equation for $\Phi(t, \B R)$ is a second-order partial
differential equation at least for two models of a random velocity
field:

{\it Model I}.  The random velocity with Gaussian statistics of
the integrals $\int _{0}^{t}\B v(t',\B {\xi}) \,d t'$ and $\int
_{0}^{t} b(t',\B {\xi}) \,d t'$, see Appendix B.

{\it Model II}.  The Gaussian velocity field with a small yet
finite correlation time, see Appendix C.

In both models equation for $\Phi(t, \B R)$ has the same form:
\begin{eqnarray} \label{WW6}
&&\partial \Phi / \partial t = \hat {\C L} \, \Phi(t,\B R) \,, \\
\nonumber
&&\hat {\C L} = B(\B R) + 2 {\B U}(\B R)\cdot \B {\nabla}
+ \hat D_{\alpha \beta}(\B R) \nabla_{\alpha} \nabla_{\beta} \,,
\end{eqnarray}
but with different expressions  for its coefficients.  The meaning
of the coefficients $ B(\B R) $, $\B U(\B R) $ and $ \hat
D_{\alpha \beta}(\B R)$ is as follows:

-- Function $ B(\B R) $ is determined only by a compressibility of the
velocity field and it causes generation of fluctuations of the number
density of inertial particles.

-- The vector $ \B U(\B R) $ determines a scale-dependent drift
velocity which describes a transfer of fluctuations of the number
density of inertial particles over the spectrum. Note that $ {\bf
U}(\B R=0) = 0 $ whereas $ B(\B R=0) \not= 0 .$ For incompressible
velocity field $ {\bf U}(\B R) = 0 ,$ $ B(\B R) = 0$.

-- The scale-dependent tensor of turbulent diffusion
$ \hat{D} _{\alpha\beta}(\B R)$ is also affected by the compressibility.

In very small scales this tensor is equal to the tensor of the
molecular (Brownian) diffusion, while in the vicinity of the
maximum scale of turbulent motions this tensor coincides with the
usual tensor of turbulent diffusion.  Tensor
$\hat{D}_{\alpha\beta}(\B R)$ may be written as
\begin{eqnarray}
\hat{D}_{\alpha\beta}(\B R) &=& 2 D \delta_{\alpha\beta} + D\sp T
_{\alpha \beta}(\B R)\,, \br D_{\alpha \beta}\sp T (\B R) &=&
\tilde{D}_{\alpha \beta}\sp T (0) - \tilde{D} _{\alpha \beta}\sp T
(\B R)\ .
\end{eqnarray}
In Appendix B we found that for Model I:
\begin{eqnarray}
 \label{eq:par}
B(\B R) &\approx& 2 \int_{0}^{\infty} \langle b[0,\B \xi(\B r_1|0)]
b[\tau,\B \xi(\B r_2|\tau)] \rangle \,d \tau\,,
\br 
{\B U}(\B R) & \approx & - 2 \int_{0}^{\infty} \langle {\B v}[0,\B
\xi(\B r_1|0)] b[\tau,\B \xi(\B r_2|\tau)] \rangle \,d\tau \,,
\br 
\tilde D_{\alpha \beta}\sp T (\B R) &\approx& 2 \int_{0}^{\infty}
\langle v_{\alpha}[0,\B \xi(\B r_1|0)] v_{\beta}[\tau,\B \xi(\B
r_2|\tau)] \rangle \,d \tau \, .
\end{eqnarray}

For the $ \delta $-correlated in time random Gaussian compressible
velocity field the operator $\hat{\C L}$ is replaced by $\hat{\C
L}_0 $ in the equation for the second moment $ \Phi(t,\B R) $,
where
\begin{eqnarray}\nn
&&\hat{\C L}_0 \equiv B_0(\B R) + 2 {\B U}_0(\B R) \cdot \B {\nabla} +
\hat D_{\alpha \beta}(\B R) \nabla_{\alpha} \nabla_{\beta} \,,
 \\ \label{W6}
&& B_0(\B R) = \nabla_{\alpha} \nabla_{\beta} \hat D_{\alpha
\beta}(\B R) \,, \br &&  U_{0,\alpha}(\B R) = \nabla_{\beta} \hat
D_{\alpha \beta}(\B R)
\end{eqnarray}
(for details see \cite{EKRS00,EKRS99}). In the $ \delta $-correlated
in time velocity field the second moment $ \Phi(t,\B R) $ can only
decay in spite of the compressibility of the velocity field. The
reason is that the differential operator $\hat{\C L}_0 \equiv
\nabla_{\alpha} \nabla_{\beta} \hat D_{\alpha \beta}(\B R) $ is
adjoint to the operator $\hat{\C L}_0^\dag \equiv \hat D_{\alpha
\beta}(\B R) \nabla_{\alpha} \nabla_{\beta}$ and their eigenvalues are
equal. The damping rate for the equation
\begin{eqnarray} \label{W1}
\partial \Phi / \partial t = \hat{\C L}_0^\dag \, \Phi(t,\B R) \,
\end{eqnarray}
has been found in Ref.~\cite{EKR95} for a compressible isotropic
homogeneous turbulence in a dissipative range:
\begin{eqnarray}
\gamma_2 = - {(3 - \sigma\sb{T})^{2} \over
6\,\tau_\eta(1 + \sigma\sb{T}) (1 + 3 \sigma\sb{T})}
\ .
\label{W4}
\end{eqnarray}
Here $\sigma\sb{T}$ is the degree of compressibility of the
tensor $D\sp T_{\alpha \beta}(\B R)$. For the $ \delta
$-correlated in time incompressible velocity field $(\sigma\sb{T}
= 0)$ Eq. (\ref{W1}) was derived in Ref.~\cite{K68}. Thus, for the
Kraichnan model of turbulent advection (with a delta correlated in
time velocity field) the clustering instability of the 2nd moment
does not occur.

A general form of the turbulent diffusion tensor in a dissipative
range is given by
\begin{eqnarray} \label{W12}
 && D^{^{\rm T}}_{\alpha \beta}(\B R)=  (C_{1} R^{2}
 \delta_{\alpha \beta} + C_{2} R_\alpha R_\beta )/\tau_ \eta \,,
\\  \nonumber
&&C_{1} = 2 (2 + \sigma\sb{T})/3\, (1 + \sigma\sb{T})\,,
\\  \nonumber
&&  C_{2} = 2 (2\sigma\sb{T} - 1)/3\,( 1 + \sigma\sb{T})\ .
\end{eqnarray}
The parameter $\sigma\sb{T}$ is defined by analogy with
Eq.~(\ref{eq:def-s}):
\begin{eqnarray}
  \label{S}
\hskip -0.4cm \sigma\sb{T}\equiv \frac{\B \nabla \cdot \B D\sb T\cdot
  \B \nabla }{\B \nabla \times \B D\sb T\times \B \nabla }=
  \frac{\nabla_\alpha \nabla_\beta D^{^{\rm T}}_{\alpha
  \beta}(\B R)}{\nabla_\alpha\nabla_\beta D^{^{\rm T}} _{\alpha'
  \beta'}(\B R) \epsilon_{\alpha\alpha'\gamma}
  \epsilon_{\beta\beta'\gamma} }\,,
\end{eqnarray}
where $ \epsilon_{\alpha\beta\gamma}$ is the fully antisymmetric
unit tensor. Equations~(\ref{eq:def-s}) and~(\ref{S}) imply that $
\sigma\sb{T}=\sigma_v$ in the case of $\delta$-correlated in time
compressible velocity field. Equations~(\ref{eq:par}) show that
for a finite correlation time identities~(\ref{W6}) are violated
and
$$  
B(\B R)\ne B_0  (\B R)\,, \quad \B U(\B R)\ne \B U_0(\B R)\ .
$$ 

For a random incompressible velocity field with a finite correlation
time the tensor of turbulent diffusion
$ D_{\alpha \beta}(\B R) = \tau^{-1} \langle \xi_\alpha(\B r_1)
\xi_\beta(\B r_2) \rangle $ [see Eq. (\ref{N1})] and the degree of
compressibility of this tensor is
\begin{eqnarray}
\label{SN}
\sigma\sb{T} = {\langle (\B \nabla \cdot \B {\xi})^{2} \rangle
\over \langle (\B \nabla {\bf \times} \B {\xi})^{2} \rangle} \,,
\end{eqnarray}
where $ \B {\xi}(\B r_1|t) $ is the Lagrangian displacement of a
particle trajectory which paths through point $ \B r_1 $ at $ t = 0 .$
Remarkably that Taylor \cite{T21} obtained the coefficient of
turbulent diffusion for the mean field in the form $ D\sb T(\B R=0) =
\tau^{-1} \langle \xi_\alpha(\B r_1|t) \xi_\alpha(\B r_1|t) \rangle$.
\subsection{Clustering instability in  Model I}
\label{mod-a}

Let us study the clustering instability for the model of the
random velocity with Gaussian statistics of the integrals
$$ 
\int  _{0}^{t}\B v(t',\B {\xi}) \,d t'\,, \quad \int _{0}^{t} b(t',\B
{\xi}) \,d t'\,,
$$ 
see Appendix B. In this model Eq.~(\ref{WW6}) in a
non-dimensional form reads:
\begin{eqnarray}
{\partial \Phi \over \partial \tilde t} &=& { \Phi'' \over m(r)} +
\Big[{1 \over m(r)} + (U-C_{2}) r^{2} \Big]{2\,\Phi' \over r } + B
\Phi \,,
\nonumber \\
1/ m(r) &\equiv & (C_{1} + C_{2}) r ^{2} + 2/{\rm Sc} \,,
\label{W14}
\end{eqnarray}
where $\B U \equiv U \B R$ and Sc$= \nu / D $ is the Schmidt
number. For small inertial particles advected by air flow Sc$ \gg
1 $. The non-dimensional variables in Eq.~(\ref{W14}) are $r
\equiv R/\eta$ and $\tilde t = t/\tau_\eta$, $B$ and $U$ are
measured in the units $\tau_\eta^{-1}$. Consider a solution of
Eq.~(\ref{W14}) in two spatial regions.

a. \it{Molecular diffusion region of scales.}  In this region $ r \ll
\text{Sc}^{-1/2}$, and all terms $\propto r^2$ (with $C_1\,,\ C_2$ and
$U$) may be neglected. Then the solution of Eq.~(\ref{W14}) is given
by
\begin{eqnarray}\nn
\Phi(r) = (1 - \alpha r^{2}) \exp(\gamma_2 t)\,,
\end{eqnarray}
where
\begin{eqnarray}\nn
\alpha &=& \text{Sc} (B - \gamma_2 \tau_\eta) / 12 \,, \quad B >
\gamma_2\tau_\eta\ .
\end{eqnarray}

b.  \it{Turbulent diffusion region of scales}. In this region
$\text{Sc}^{-1/2} \ll r \ll 1$, the molecular diffusion term $\propto
1/\text{Sc}$ is negligible. Thus, the solution of~(\ref{W14}) in
this region is
\begin{eqnarray}\nn
\Phi(r) = A_{1} r ^{-\lambda} \exp(\gamma_2 t)\,,
\end{eqnarray}
where
\begin{eqnarray}\nn
\lambda &=& (C_{1} - C_{2} + 2 U \pm i
C_3) / 2 (C_{1} + C_{2})\,,  \br 
C_3^2&=&4 (B-\gamma_2\tau_\eta)(C_1+C_2) - (C_{1} - C_{2} + 2
U)^{2} \ .
\end{eqnarray}
Since the total number of particles in a closed volume is
conserved:
$$ 
\int_{0}^{\infty} r^{2} \Phi(r) \,d r = 0\ .
$$ 
 This implies that $ C_3^2 > 0$, and therefore $ \lambda $ is a
 complex number. Since the correlation function $ \Phi(r) $ has a
 global maximum at $ r = 0 ,$ $ \, C_{1} > C_{2}- 2U$. The latter
 condition for very small $U$ yields $ \sigma_{T} \leq 3$. For $r \gg
 1$ the solution for $\Phi(r)$ decays sharply with $r$. The growth
 rate $ \gamma_2$ of the second moment of particles number density can
 be obtained by matching the correlation function $ \Phi(r) $ and its
 first derivative $ \Phi'(r) $ at the boundaries of the above regions,
 i.e., at the points $r = \text{Sc}^{-1/2}$ and $r = 1$. The matching
 yields $ C_3 / 2(C_{1} + C _{2}) \approx 2 \pi / \ln \text{Sc}
 $. Thus,
\begin{eqnarray}\nn
\gamma_2 &=& {1 \over \tau_\eta (1 + 3 \sigma\sb{T})}
\biggl[{200 \sigma\sb{U} (\sigma\sb{T} - \sigma\sb{U})
\over 3 (1 + \sigma\sb{U})} - {(3 - \sigma\sb{T})^{2} \over 6 (1 +
\sigma\sb{T})}
\\ \label{W17}
&&\hskip -1cm - {3 \pi^2 (1 + 3 \sigma\sb{T})^{2}
\over (1 + \sigma\sb{T}) \ln^2 \text{Sc}} \biggr]
+ {20 (\sigma\sb{B} - \sigma\sb{U})
\over \tau_\eta (1 + \sigma\sb{B}) (1 + \sigma\sb{U})} \,,
\end{eqnarray}
where we introduced parameters $\sigma\sb{B}$ and $\sigma\sb{U}$
defined by
\begin{equation}
  \label{eq:defs}
   B = 20 \sigma\sb{B} / (1 + \sigma\sb{B})\,, \quad
U = 20 \sigma\sb{U} / 3 (1 + \sigma\sb{U})\ .
\end{equation}
Note that the parameters $ \sigma\sb{B} \approx \sigma\sb{U} \sim
\sigma_v $. For the $ \delta $-correlated in time random
compressible velocity field $ \sigma\sb{B} = \sigma\sb{U} =
\sigma\sb{T} = \sigma_v $. Figure 1 shows the range of parameters
$ (\sigma_v , \sigma\sb{T}) $ for $ \sigma\sb{B} = \sigma\sb{U} =
\sigma_v $ in the case of $ \text{Sc} = 10^3 $ (curve c), $
\text{Sc} = 10^5 $ (curve b) and $ \text{Sc} \to \infty $ (curve
a). The dashed line $ \sigma_v = \sigma\sb{T} $ corresponds to the
$ \delta $-correlated in time random compressible velocity field.
This is a limiting line for the curve "a". Figure 1 demonstrates
that even a very small deviations from the $ \delta $-correlated
in time random compressible velocity field results in the
instability of the second moment of the number density of inertial
particles. The minimum value of $ \sigma\sb{T} $ required for the
clustering instability is $ \sigma\sb{T} \approx 0.26 $ and a
corresponding value of $\sigma_v \approx 0.12 $ (see Fig. 1). For
smaller value of $\sigma_v$ the clustering instability can occur,
but it requires larger values of $ \sigma\sb{T} .$

\begin{figure}
\includegraphics[width=8cm]{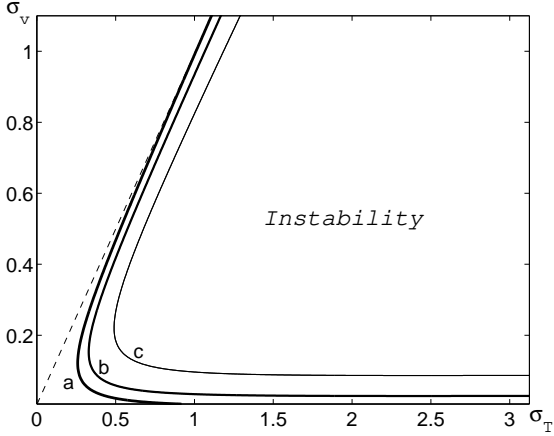}
\caption{\label{Fig1} The range of parameters $ (\sigma_v ,
\sigma\sb{T}) $ for $ \sigma\sb{B} = \sigma\sb{U} = \sigma_v $ in
the case of $ \text{Sc} = 10^3 $ (curve c), $ \text{Sc} = 10^5 $
(curve b) and $ \text{Sc} \to \infty $ (curve a). The dashed line
$ \sigma_v = \sigma\sb{T} $ corresponds to the $ \delta
$-correlated in time random compressible velocity field.}
\end{figure}

Notably, in Model II of a random velocity field (\ie, the Gaussian
velocity field with a small yet finite correlation time) the
clustering instability occurs when $\sigma_v > 0.2$ (see Appendix
C). Indeed, the growth rate $ \gamma_2$ of the second moment of
particles number density is determined by equation:
\begin{eqnarray}  \nn
\Gamma &=& \tilde B(\sigma_v) {\rm St}^{2} - \frac{(3 -
\sigma_v)^{2} }{6 (1 + \sigma_v) (1 + 3 \sigma_v)}  \br && - {8 (1
+ 3 \sigma_v) \over 3 (1 + \sigma_v)} \biggl({\pi \over \ln {\rm
Sc}} \biggr)^{2} \;,  \br \tilde B(\sigma_v) &=& 12 \left(b_{2} +
\frac{b_{3} a_{1}}{4 a_{2}^{2}} - \frac{b_{1}}{2 a_{2}} \right)\
\; ,    \label{N20}
\end{eqnarray}
where St$ = \bar\tau\sb{ren} / \tau_{\eta} $ is the Strouhal
number, $ \Gamma = \gamma_2 (1 + \bar\tau\sb{ren} \gamma_2)^{2} ,$
and
\begin{eqnarray*}
a_{1} &=& {2(19 \sigma_v + 3) \over 3(1 + \sigma_v)} \;,   \qquad
\qquad a_{2} = {2(3 \sigma_v + 1) \over 3(1 + \sigma_v)} \;,
\\
b_{1} &=& - {1 \over 27 (1 + \sigma_v)^{2}} (12 - 1278 \sigma_v -
3067 \sigma_v^{2}) \;,
\\
b_{2} &=& {850 \over 9} \biggl({\sigma_v \over 1 + \sigma_v}
\biggr)^{2} \;,
\\
b_{3} &=& {1 \over 27 (1 + \sigma_v)^{2}} (36 + 466 \sigma_v +
2499 \sigma_v^{2}) \ .
\end{eqnarray*}
Figure 2 shows the range of parameters $ (\text{Sc}, \sigma_v) $
for $ \text{Sc} = 10^3 $ (curve c), $ \text{Sc} = 10^5 $ (curve b)
and $ \text{Sc} \to \infty $ (curve a).

\begin{figure}
\includegraphics[width=8cm]{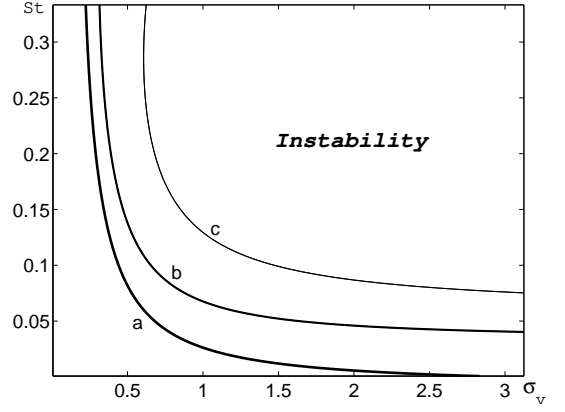}
\caption{\label{Fig2} The range of parameters $ (\text{Sc},
\sigma_v) $ for $ \text{Sc} = 10^3 $ (curve c), $ \text{Sc} = 10^5
$ (curve b) and $ \text{Sc} \to \infty $ (curve a). The line $
\text{Sc} =0 $ corresponds to the $ \delta $-correlated in time
random compressible velocity field.}
\end{figure}

It is seen from FIG.~\ref{Fig2} that for $ \sigma_v > 0.2 $ the
second-order correlation function of the number density of
inertial particles can grow in time exponentially (i.e., $
\gamma_2 > 0) $ even for very small Strouhal numbers. For example,
in the vicinity of $ \sigma_v = 3 ,$ the growth rate $ \gamma_2 $
of the clustering instability of the second-order correlation
function is given by
\begin{equation}
\gamma_2 =4 \times 10^{3} \biggl({\tau u_{\eta} \over \eta}
\biggr)^{2} - {(3 - \sigma_v)^{2} \over 240} - {20 \over 3}
\biggl({\pi \over \ln {\rm Sc}} \biggr)^{2} \; . \label{N23}
\end{equation}
The sufficient condition for the exponential growth of the second
moment of a number density of inertial particles is $ {\rm Sc} >
{\rm Sc}^{({\rm cr})} ,$ where the critical Schmidt number $ {\rm
Sc}^{({\rm cr})} $ is given by $ {\rm Sc}^{({\rm cr})} = {\rm
Sc}(\gamma_2=0) .$ The clustering instability occurs when the
degree of compressibility of particles velocity $ \sigma_v
> 0.2 ,$ \ie, for particles and droplets with the radius $
a_{\ast} > 25.4 \, \mu$m\,. Equation~(\ref{W27}) also yields a
similar value $\sigma_{\rm cr}\sim 1/6$ for the threshold of the
instability of the 2nd moment (at $q_{\rm cr} =2$). Note that
Eq.~(\ref{W4}) is written for $\text{Sc} \to \infty$.

\section{Nonlinear effects}
\label{nonlin}

The compressibility of the turbulent velocity field with a finite
correlation time can cause the exponential growth of the moments of
particles number density. This small-scale instability results in
formation of strong inhomogeneities (clusters) in particles spatial
distributions. The linear analysis does not allow to determine a
mechanism of saturation of the clustering instability. As can be seen
from Eq.~(\ref{W17}) molecular diffusion only depletes the growth
rates of the clustering instability at the linear stage (contrary to
the instability discussed in Ref. \cite{BF01}). The clustering
instability is saturated by nonlinear effects.

Now let us discuss a mechanism of the nonlinear saturation of the
clustering instability using on the example of atmospheric
turbulence with characteristic parameters: $\eta\sim 1$mm,
$\tau_\eta \sim (0.1 - 0.01)$s.  A momentum coupling of particles
and turbulent fluid is essential when $ m_{\rm p} n\sb{cl} \sim
\rho$, i.e., the mass loading parameter $\phi= m_{\rm p} n\sb{cl}
/ \rho $ is of the order of unity (see, e.g., \cite{CST98}). This
condition implies that the kinetic energy of fluid $ \rho \langle
{\B u}^{2} \rangle $ is of the order of the particles kinetic
energy $ m_{\rm p} n\sb{cl} \langle {\B v}^{2} \rangle ,$ where $
|{\B u}| \sim |{\B v}|$. This yields:
\begin{eqnarray}
  \label{eq:est}
   n\sb{cl} \sim a^{-3} (\rho/3 \rho_{\rm p}) \ .
\end{eqnarray}
For water droplets $ \rho_{\rm p} / \rho \sim 10^{3}$. Thus, for $
a = a_* \sim 30 \mu$m we obtain $ n\sb{cl} \sim 10^{4} $ cm$^{-3}$
and the total number of particles in the cluster of size $\eta$,
$N_{\rm cl}\simeq \eta^3 n\sb{cl} \sim 10$. This values may be
considered as a lower estimate for the ``two-way coupling'' when
the effect of fluid on particles has to be considered together
with the feed-back effect of the particles on the carrier fluid.
However, it was found in~\cite{02LP} that turbulence modification
by particles is governed by the ratio of the particle energy and
\it{ the total energy of the suspension} (rather then the energy
of the carrier fluid) and thus by parameter $\phi(1+\phi)$ (rather
then by $\phi$ itself). Thus we expect that the two-way coupling
can only mitigate but not stop the clustering instability.

An actual mechanism of the nonlinear saturation of the clustering
instability is ``four way coupling'' when the particle-particle
interaction is also important. In this situation the particles
collisions result in effective particle pressure which prevents
further grows of concentration. Particles collisions play
essential role when during the life-time of a cluster the total
number of collisions is of the order of number of particles in the
cluster. The rate of collisions $ J \sim n\sb{cl} / \tau_\eta$ can
be estimated as $ J \sim 4 \pi a^{2} n\sb{cl}^{2} |{\B v}_{\rm
rel}|$. The relative velocity ${\B v}_{\rm rel}$ of colliding
particles with different but comparable sizes can be estimated as
$ |{\B v}_{\rm rel}| \sim \tau_{\rm p} | ({\B u} \cdot \B
{\nabla}) {\B u}| \sim \tau_{\rm p} u_{\eta}^{2} / \eta$. Thus the
collisions in clusters may be essential for
\begin{eqnarray}
  \label{eq:nc}
   n\sb{cl} \sim a^{-3} (\eta/a)  (\rho/3 \rho_{\rm p}) \,,\quad
   \ell_{\rm s}  \sim a (3 a \rho_{\rm p}/\eta \rho)^{1/3}\,,
\end{eqnarray}
 where $\ell_{\rm s}$ is a mean separation of particles in the
 cluster.  For the above parameters ($a=30\mu$m) $ n\sb{cl} \sim
 3\times 10^{5}$cm$^{-3}$, $ \ell_{\rm s} \sim 5 \,a\approx 150 \mu$m
 and $ N\sb{cl}\sim 300$. Note that the mean number density of
 droplets in clouds $\bar n$ is about $10^2 - 10
 ^3$cm$^{-3}$. Therefore the {\em clustering instability of droplets
 in the clouds increases their concentrations in the clusters by the
 orders of magnitude}.

In all our analysis we have neglected the effect of sedimentation
of particles in gravity field which is essential for particles of
the radius $ a > 100 \mu$m. Taking $\ell\sb{cl}\simeq \eta$ we
assumed implicitly that $\tau_{\rm p}< \tau_\eta$. This is valid
(for the atmospheric conditions) if $ a \leq 60 \mu$m.  Otherwise
the cluster size can be estimated as $\ell\sb{cl}\simeq \eta
(\tau_{\rm p}/\tau_\eta)^{3/2}$.

Our estimates support the conjecture that {\em the clustering
instability serves as a preliminary stage for a coagulation of water
droplets in clouds leading to a rain formation}.

\section{Discussion}
\label{discus}

In this study we investigated the clustering instability of the
spatial distribution of inertial particles advected by a turbulent
velocity field. The instability results in formation of clusters,
{\em i.e.,} small-scale inhomogeneities of aerosols and droplets.
The clustering instability is caused by a combined effect of the
particle inertia and finite correlation time of the velocity
field. The finite correlation time of the turbulent velocity field
causes the compressibility of the field of Lagrangian
trajectories. The latter implies that the number of particles
flowing into a small control volume in a Lagrangian frame does not
equal to the number of particles flowing out from this control
volume during a correlation time. This can result in the depletion
of turbulent diffusion.

The role of the compressibility of the velocity field is as
follows. Divergence of the velocity field of the inertial
particles $ \text{div}\,\B v = \tau_{p} \Delta P / \rho .$ The
inertia of particles results in that particles inside the
turbulent eddies are carried out to the boundary regions between
the eddies by inertial forces (i.e., regions with low vorticity
and high strain rate).  For a small molecular diffusivity $
\text{div}\,\B v \propto - dn / dt $ [see Eq. (\ref{eq:T1})].
Therefore, $ dn / dt \propto - \tau_p \Delta P / \rho .$ Thus
there is accumulation of inertial particles (i.e., $ dn / dt > 0 )
$ in regions with $ \Delta P < 0 . $ Similarly, there is an
outflow of inertial particles from the regions with $ \Delta P > 0
.$ This mechanism acts in a wide range of scales of a turbulent
fluid flow. Turbulent diffusion results in relaxation of
fluctuations of particles concentration in large scales. However,
in small scales where turbulent diffusion is small, the relaxation
of fluctuations of particle concentration is very weak.  Therefore
the fluctuations of particle concentration are localized in the
small scales.

This phenomenon is considered for the case when density of fluid is
much less than the material density $ \rho_p $ of particles $ (\rho
\ll \rho_p). $ When $ \rho \geq \rho_p $ the results coincide with
those obtained for the case $ \rho \ll \rho_p $ except for the
transformation $ \tau_p \to \beta_{\ast} \tau_p , $ where
\begin{eqnarray*}
\beta_{\ast} = 2 \biggl(1 + {\rho \over \rho_p} \biggr)
\biggl({\rho_p - \rho \over 2\rho_p + \rho} \biggr) \; .
\end{eqnarray*}
For $ \rho \geq \rho_p $ the value $ dn / dt \propto - \beta_{\ast}
\tau_p \Delta P / \rho .$ Thus there is accumulation of inertial
particles (i.e., $ dn / dt > 0 ) $ in regions with the minimum
pressure of a turbulent fluid since $ \beta_{\ast} < 0 .$ In the case
$ \rho \geq \rho_p $ we used the equation of motion of particles in
fluid flow which takes into account contributions due to the pressure
gradient in the fluid surrounding the particle (caused by acceleration
of the fluid) and the virtual ("added") mass of the particles relative
to the ambient fluid \cite{MR83}.

The exponential growth of the second moment of a number density of
inertial particles due to the small-scale instability can be
saturated by the nonlinear effects (see Section IV). The
excitation of the second moment of a number density of particles
requires two kinds of compressibilities: compressibility of the
velocity field and compressibility of the field of Lagrangian
trajectories, which is caused by a finite correlation time of a
random velocity field. Remarkably, the compressibility of the
field of Lagrangian trajectories determines the coefficient of
turbulent diffusion (i.e. the coefficient $D_{\alpha \beta}$ near
the second-order spatial derivative of the second moment of a
number density of inertial particles in Eq.  (\ref{WW6}). The
compressibility of the field of Lagrangian trajectories causes
depletion of turbulent diffusion in small scales even for
$\sigma_{v} = 0$.  On the other hand, the compressibility of the
velocity field determines a coefficient B(r) near the second
moment of a number density of inertial particles in Eq.
(\ref{WW6}). This term is responsible for the exponential growth
of the second moment of a number density of particles.

\vskip 0.2cm

{\bf Summary}:

$\bullet$ We showed that the physical reason for the {\em
clustering instability} in spatial distribution of particles in
turbulent flows is a combined effect of the inertia of particles
leading to a compressibility of the particle velocity field $\B
v(t,\B r)$ and a finite velocity correlation time.

$\bullet$ The clustering instability can result in a {\em strong
clustering} whereby a finite fraction of particles is accumulated
in the clusters and a {\em weak clustering} when a finite fraction
of particle collisions occurs in the clusters.

$\bullet$ The crucial parameter for the clustering instability is a
radius of the particles $a$. The instability criterion is $a> a_{\rm
cr} \approx a_*$ for which $\langle (\text{div}\, \B v)^2 \rangle
=\langle |\text{rot} \, \B v|^2\rangle$. For the droplets in the
atmosphere $a_*\simeq 30\mu$m. The growth rate of the clustering
instability $\gamma\sb{cl} \sim \tau_\eta ^{-1} (a/a_*)^4$, where
$\tau_\eta$ is the turnover time in the viscous scales of turbulence.

$\bullet$ We introduced a new concept of compressibility of the
turbulent diffusion tensor caused by a finite correlation time of
an incompressible velocity field. For this model of the velocity
field, the field of Lagrangian trajectories is not
divergence-free.

$\bullet$ We suggested a mechanism of saturation of the clustering
instability - {\em particle collisions in the clusters}.  An evaluated
nonlinear level of the saturation of the droplets number density in
clouds exceeds by the orders of magnitude their mean number density.

\begin{acknowledgments}
We have benefited from discussions with I. Procaccia. This work
was partially supported by the German-Israeli Project Cooperation
(DIP) administrated by the Federal Ministry of Education and
Research (BMBF), by the Israel Science Foundation and by INTAS
(Grant 00-0309). DS is grateful to a special fund for visiting
senior scientists of the Faculty of Engineering of the Ben-Gurion
University of the Negev and to the Russian Foundation for Basic
Research (RFBR) for financial support under grant 01-02-16158.
\end{acknowledgments}

\appendix
  \S{A}{B\lowercase {asic equations in the model with a random renewal
  time}}

In this Appendix we derive \REF{C14} for the simultaneous
second-order correlation function $\Phi\tr$ which serves as a
basis for further analysis in Appendixes~\ref{s:modelB} and
\ref{s:modelC} under some simplifying model assumptions about the
statistics of the velocity field.

\SS{method}{Exact solution of dynamical equations for a given
velocity field} 
\SSS{no-dif}{Simple case: no molecular diffusion} 
Consider first \REF{T1} for the number density of particles $n\tr$ in
the case $D = 0$:
\begin{eqnarray}
{\partial n (t,\B r ) \over \partial t}
+ \B \nabla \cdot[n\tr \B v\tr] = 0\,,
\label{eq:A1}
\end{eqnarray}
when all particles are transported only by advection.  Solution of
\REF{A1} with the initial condition $n(s,\B r)$ is given by
\begin{eqnarray}   \label{eq:A2}
n(t,\B r) &=& G(t,\B r) \, n[s, \B \xi \sb L(t,\r|s)] \,,
\end{eqnarray}
where $\B \xi\sb L (t,\r| s)$ is the Lagrangian trajectory of the
particle which is located at coordinate $\r$ at time $t$. Here we
label the particles at \it{present} moment of time $t$ and
consider a \it{current} time $s<t$ as moments in the past. This
differs from a usual approach, see Eqs.~\Ref{rL}, when particles
are labelled at the \it{initial} time $t_0$, and a current time
$t>t_0$. Therefore in the equations below it is more convenient to
redefine Lagrangian displacement $\B \rho\sb L(t,r|s)\to \tilde{\B
\rho}\sb L(t,r|s) =- \B \rho\sb L(t,r|s)$. Now Eqs.~\Ref{rL} can
be written as
\begin{eqnarray}
\label{eq:rL1} \tilde{\B \rho}\sb L (t,\r |s) &=& \int^{t}_{s} \B
v [\tau,\B \xi \sb L (t,\r | \tau)] \,\d \tau \,,
\\ \label{eq:xiL1} 
\B \xi \sb L (t,\r |s) &\equiv &\B r-\tilde{\B \rho}\sb L(t,\r |
s)\ .
\end{eqnarray}
The Green function is the functional of $\B \xi\sb L (t,\r | s)$:
\begin{eqnarray}\nn
 G(t,\r,s) &=& \exp \Big\{- \int_{s}^{t} b[\tau,\B \xi\sb L (t,\r |
 \tau )]\,d \tau\Big\} \,,\\ \label{eq:green} b(t,\B r) &\equiv& \B
 \nabla \cdot \B v\tr \,,
\end{eqnarray}
Introduce the \it{shift operator}
\begin{eqnarray}
  \label{eq:shift}
   \exp[- \tilde{\B \rho}\sb L \cdot \B {\nabla}\,] = 1 - \tilde{\B
   \rho}\sb L \cdot \B {\nabla} + \frac{1}{2!} [- \tilde{\B \rho}\sb L
   \cdot \B {\nabla}]^{2} - \dots
\end{eqnarray}
which acts as follows:
\begin{eqnarray}
  \label{eq:shift1}
   \exp[- \tilde{\B \rho}\sb L \cdot \B
   {\nabla}\,]n\tr=n(t,\r-\tilde{\B \rho}\sb L)\ .
\end{eqnarray}
One can validate relation \Ref{shift1} by Taylor series expansion
of the function $ n(t,\r - \tilde{\B \rho}\sb L)$. Now \REF{A2}
can be rewritten as follows:
\begin{eqnarray}
n(t,\r) = G(t,\r,s) \, \exp[- \tilde{\B \rho}\sb L (t,\r | s) \cdot
\B {\nabla}\,]
n(s,\r) \; .
\label{eq:A4}
\end{eqnarray}
\SSS{wiener}{Molecular diffusion as a Wiener process}
Consider now the full \REF{T1} with $ D \ne 0$ whereby particles
are transported by both, fluid advection and molecular diffusion.
It was found by Wiener (see, e.g., \cite{ZRS90}) that Brownian
motion (molecular diffusion) can be described by the \it { Wiener}
random process $\B w(t)$ with the following properties:
\begin{eqnarray}\label{eq:W}
\langle {\B w(t)} \rangle_{\B w}=0\,,\quad \langle w_i(t+\tau)
w_j(t) \rangle_{\B w}= \tau \delta _{ij}\ .
\end{eqnarray}
Here $ \langle \dots \rangle_{\B w} $ denotes the mathematical
expectation over the statistics of the Wiener process. Introduce
the Wiener trajectory $\B \xi\sb W(t,\r|s)$ (which usually is
called the \it{Wiener path}) and the Wiener displacement $\B
\rho\sb W(t,\r|s)$ as follows:
\begin{eqnarray}
\label{eq:rW}
\B \xi _{\B w} (t,\r |s) &\equiv &\B r-\B \rho\sb W(t,\r | s)\,,
\br 
\B \rho\sb W(t,\r |s) &=& \int^{t}_{s} \B v [\tau,\B \xi_{\B w}
(t,\r | \tau)] \,\d \tau + \sqrt{2 D} {\B w}(t-s) \ .
\end{eqnarray}
Comparison of this formula with Eqs.~\Ref{rL1} shows that in the
limit $D\to 0$,  $\B \xi \sb W (t,\r |s)\to \B \xi \sb L (t,\r
|s)$ and $\B \rho \sb W (t,\r |s)\to \B \rho \sb L (t,\r |s)$.

In Refs.~[\onlinecite{EKR00}] it was shown that solution of
\REF{T1} (with $D\ne 0$) can be written as solution \Ref{A4} of
\REF{A1} (with $D= 0$) by replacement $\tilde{\B \rho}\sb L
(t,\r|s)\to \tilde{\B \rho}\sb W (t,\r|s)$ and then averaging over
the statistics of the Wiener processes~\Ref{W}:
\begin{eqnarray}\nn
\hskip -0.1cm n(t,\r)=\<  G(t,\r,s) \, \exp[-\B \rho\sb W (t,\r |
s) \cdot \B {\nabla}\,] n(s,\r) \>_{\B w}\ .  \\
\label{eq:A5}
\end{eqnarray}

\SS{velocity}{Two-step averaging over velocity statistics}
\SSS{modelA}{Model of a random velocity field}

Note that \REF{A5} is a solution of \REF{T1} at a \it{given}
realization of the random velocity field. Our next goal is to
determine the simultaneous correlation functions
\begin{eqnarray}
  \label{eq:def-cor}
  \bar n(t) &=& \lla  n(t,\r) \rra_{\B v}\,,\br
 \Phi(t,\r_2-\r_1) &=& \lla
n(t,\r_1) n(t,\r_2) \rra_{\B v}- \bar
 n^2(t)\,,
\end{eqnarray}
averaged over the stationary, space homogeneous statistics of
turbulent velocity field, where $\lla \dots \rra_{\B v}$ denotes
this averaging. Since the initial distribution $n(t_0,\r)$ is
assumed to be homogeneous in space, $\bar n(t)$ is independent of
spatial coordinate, and $ \Phi(t,\r_2-\r_1)$ depends only on the
difference $\r_2-\r_1$.

In order to simplify the averaging procedure \Ref{def-cor} we
consider a model of random velocity field which fully looses
memory at some instants of renewal $\tau_j$.  For $t_1$ and $t_2$
inside a renewal interval [$\tau_j<t_1,t_2<\tau_{j+1}$] the
velocity pair correlation function is defined as
\begin{eqnarray}
  \label{eq:def-CF}
  \C F^{\alpha\beta}(t_2-t_1,\r_2-\r_1)\equiv \la
  v_\alpha(t_1,\r_1)v_\beta(t_2,\r_2)\ra_{\B v}\,,
\end{eqnarray}
where $\la\dots\ra_{\B v}$ denotes averaging over ``intrinsic
statistics'' of the velocity field. In our model the velocity
fields before and after renewals are statistically independent.
The interval between the renewal instants $\tau_j$ may be the same
or randomly distributed, say with the Poisson statistics. In the
latter case the full averaging $\lla \dots \rra_{\B v}$ may be
considered as a two-stage process. First one calculates $\la \dots
\ra_{\B v}$ and then averages over the statistics of the renewal
time $\tau\sb{ren}$, which is denoted as $\la \dots \ra\sb{ren}$:
\begin{eqnarray}
  \label{eq:full-ev}
  \la \!\la \dots \ra \! \ra_{\B v}
\equiv\la \ \la \dots \ra_{\B v}\ra\sb{ren}\ .
\end{eqnarray}
For the Poisson statistics of $\tau_j$
\begin{eqnarray}\nn
 &&
F^{\alpha\beta}(t_2-t_1,\r_2-\r_1)\equiv \la \!\la
  v_\alpha(t_1,\r_1)v_\beta(t_2,\r_2)\ra \! \ra_{\B v}\\  \label{eq:FF}
\hskip -1cm &=& \C F^{\alpha\beta}(t_2-t_1,\r_2-\r_1) \exp(-  |t_2-t_1|
/
\bar\tau\sb{ren})\,,
\end{eqnarray}
where $\bar\tau\sb{ren}$ is a mean renewal time. It would be
useful to define the correlation time of the function $\C
F^{\alpha\beta}$ as follows
\begin{eqnarray}
\tau_{v}(R) = \int \C F^{\alpha\beta}(\tau,\B R)  \,d \tau /
\C F^{\alpha\beta}(0,\B R) \  .
\label{C10}
\end{eqnarray}
Certainly this model of the random velocity field cannot be
considered as universal. However, it reproduces important features
of some flows (see, e.g., Ref.~[\onlinecite{LST00}]).

\SSS{twostep}{Averaging procedure} Our model involves three random
processes:
\begin{enumerate}

\item The Wiener random process which describes Brownian
(molecular) diffusion.

\item  Poisson process for a random renewal time.

\item  The random velocity field between the renewals.
\end{enumerate}
\REF{A5} presents $n\tr$ after the first step, i.e., it describes
the number density at a \it{given} realization of a velocity
field. Using \REF{A5} we obtain
\begin{eqnarray}
&& n(t,\r_1) n(t,\r_2) = \la G(\r_1) G(\r_2) \,
\exp[\B {\xi}'(\r_1) \cdot \B {\nabla}_{1}
\nonumber \\
&& + \B {\xi}'(\r_2) \cdot \B {\nabla}_{2}] n(s,\r_1)
n(s,\r_2) \ra_{\B w\B w}\,,
\label{eq:AA3A}
\end{eqnarray}
where $ \B {\nabla}_{1} = \partial / \partial \r_1 $ and $ \B
{\nabla}_{2} = \partial / \partial \r_2$ and $\la \ra_{\B w\B w}$
denotes averaging over two independent Wiener processes
determining two Wiener paths.  Hereafter for simplicity we use the
following notations: $ G(\r) \equiv G(t,\r,s) $ and $ \B
{\xi}'(\r) \equiv \B {\xi}'(t,\r | s).$

Now we average Eq. (\ref{eq:AA3A}) over a random velocity field
for a given realization of a Poisson process:
\begin{eqnarray} \label{eq:C11}
&& \tilde \Phi(t,\B r_2-\B r_1)  =
 \langle n(t,\B r_1) n(t,\B r_2)
\rangle_{\B v} - (\bar n)^{2} \br
 &=& \langle \langle G(\B r_1) G(\B r_2) \,
\exp[\B {\xi}'(\B r_1) \cdot \B {\nabla}_{1}
+ \B {\xi}'(\B r_2)
\cdot \B {\nabla}_{2}] \rangle_{\B w\B w}  \rangle_{\B v} \br
  && \times \tilde \Phi(t_{0},\B r_1-\B r_2) \} \ .
\end{eqnarray}
Here the time $ t_{0} $ is the last renewal time before time $ t $
and $ t' = t - t_{0} $ is a random variable.  Thus, averaging of
the functions
$$ 
G(\B r_1) G(\B r_2) \, \exp[\B {\xi}'(\B r_1)
\cdot \B {\nabla}_{1} + \B {\xi}'(\B r_2) \cdot \B {\nabla}_{2}]\,,
 \Phi(t_{0},\B r_1-\B r_2)
$$ 
is decoupled into two time intervals because the first function is
determined by the velocity field after the renewal while the second
function $ \Phi(t_{0},\B r_1-\B r_2) $ is determined by the velocity
field before renewal.  Now we take into account that for the Poisson
process any instant can be chosen as the initial instant. We average
\REF{C11} over the random renewal time. The probability density $
p(t') $ for a random renewal time is given by
\begin{eqnarray}
p(t') = \bar\tau^{-1} \sb{ren} \exp(- t' / \bar\tau\sb{ren}) \; .
\label{eq:C12}
\end{eqnarray}
Thus the resulting averaged equation for ``fully'' averaged
correlation function  $\Phi(t,\B R)=\langle \tilde \Phi(t,\B R)
\rangle\sb{ren}$, defined by \REF{def-cor}, assumes the following
form:
\begin{eqnarray}
&& \hskip -0.4cm \Phi(t,\B R) = \bar\tau^{-1}\sb{ren}
\int_{0}^{t} \hat P(\tau,\B R)
\Phi(t-\tau,\B R)\exp(- \tau / \bar\tau\sb{ren}) \, d \tau \nonumber \\
&&\hskip 1cm + \exp(- t / \bar\tau\sb{ren}) \hat P(t,\B R)
\Phi_{0}(\B R) \, . \label{eq:C14}
\end{eqnarray}
The first term in \REF{C14} describes the case when there is at
least one renewal of the velocity field during the time $ t $
(i.e.,  the Poisson event), whereas the second term describes the
case when there is no renewal during the time $ t$. Here $
\Phi_{0}(\B R) = \Phi(t=0,\B R) $ and
\begin{eqnarray} \label{C2}
&&\hat P(t,\B R) = \langle \langle G(\B r_1)
G(\B r_2) \, \br
&&
\hskip 0.5cm \times \exp[\B {\xi}'(\B r_1) \cdot \B {\nabla}_{1}
+ \B {\xi}'(\B r_2) \cdot \B {\nabla}_{2}] \rangle_{\B w\B w}
\rangle_{\B v} \br 
&& \hskip -0.4cm
=\exp \lla g(\B r_1) + g(\B r_2)+\B {\xi}'(\B
r_1) \cdot \B {\nabla}_{1} + \B {\xi}'(\B r_2) \cdot \B {\nabla}_{2}
\rangle_{\B w\B w} \rangle_{\B v} \,,
\end{eqnarray}
where $ G(\B r) = \exp[g(\B r)] $.  Equation~\Ref{C14} is
simplified in Appendixes B and C under the additional assumptions
about the velocity field statistics.

\S{modelB}{velocity field with Gaussian Lagrangian trajectories}
Consider a model of a random velocity field where Lagrangian
trajectories, i.e., \ \it{the integrals $ \int \B v (\mu,\B {\xi})
\,d \mu $ and $ \int b(\mu,\B {\xi}) \,d \mu $} have Gaussian
statistics. Using an identity $ \langle \exp(a \eta)
\rangle_{\eta} = \exp(\hf a^2) $ in Eq. (\ref{C2}) we obtain
\begin{eqnarray}\label{eq:C15}
\hat P(\mu,\B R) &=&\exp [\mu \hat {\cal L}] \;,
\end{eqnarray}
where
\begin{eqnarray}\nn
\hat {\cal L} &=& B(\B R) + 2 U_{\alpha}(\B R) \nabla_{\alpha} +
\hat D_{\alpha \beta}(\B R) \nabla_{\alpha} \nabla_{\beta} \,,
\\
\label{eq:C16}
\mu B(\B R) &=& \lla  g(\B r_1) g(\B r_2)
\rangle_{\B w\B w} \rangle_{\B v} \;, \br
\mu U_{\alpha}(\B R) &=&
- \lla \xi'_{\alpha}(\B r_1) g(\B r_2) \rangle_{\B w\B w}
\rangle_{\B v} \;, \br
\hat D_{\alpha \beta}(\B R) &=& D_{\alpha
\beta}(0) - D_{\alpha \beta}(\B R) \;, \br
\mu D_{\alpha \beta}(\B
R) &=& \lla \xi'_{\alpha}(\B r_1) \xi'_{\beta}(\B r_2) \rangle_{\B
w\B w} \rangle_{\B v} \ .
\label{C17}
\end{eqnarray}
Here $ \eta $ is a Gaussian random variable with zero mean value
and unit variance and $ \lla G(\B r) \rangle_{\B w} \rangle_{\B v}
= 1 .$ The latter yields
$$ 
 \lla g \rangle_{\B w} \rangle_{\B v} = - \hf \lla \tilde g^{2}
 \rangle_{\B w} \rangle_{\B v}\,, \quad g = \lla g \rangle_{\B w}
 \rangle_{\B v} + \tilde g\,,
$$ 
 where $ \langle \langle \tilde g \rangle_{\B w} \rangle_{\B v} = 0 .$
 When correlation time $ \tau_{v}(R)$, Eq.~(\ref{C10}), is much less
 then the current time $t $ and $\bar\tau\sb{ren}$, these correlation
 functions are given by
\begin{eqnarray}
B(\B R) &\!\!\!=& \!\! \! 2 \!\!  \int\limits_{0}^{\infty}
\lla b[0,\B {\xi}(\B r_1)]
b[\mu',\B {\xi}(\B r_2)] \rangle_{\B w\B w} \rangle_{\B v}
\,d \mu' ,
\label{eq:C18} \br 
U_{\alpha}(\B R) &\!\!\!=& \!\! \! - 2 \!\!
\int\limits_{0}^{\infty} \lla v_{\alpha}[0,\B {\xi}(\B r_1)]
b[\mu',\B {\xi}(\B r_2)] \rangle_{\B w\B w} \rangle_{\B v} \,d
\mu'
\;, \br 
D_{\alpha \beta}(\B R) &\!\!\!=& \!\! \! 2 \!\!
\int\limits_{0}^{\infty} \lla v_{\alpha}[0,\B {\xi}(\B r_1)]
v_{\beta}[\mu',\B {\xi}(\B r_2)] \rangle_{\B w\B w} \rangle_{\B v}
\,d \mu' \;,
\label{C20}
\end{eqnarray}
where we used an identity
\begin{eqnarray*}
&& \lla \int\limits_{0}^\mu a_{\alpha}(\mu',\B r_1) \,d
\mu' \int\limits_{0}^{\mu} c_{\beta}(\mu'',\B r_2) \,d \mu''
\rangle_{\B w} \rangle_{\B v}
\\  
&&  \hskip 0.5cm \simeq 2 \mu \int\limits_{0}^{\infty} \langle \langle
a_{\alpha}(0,\B r_1) c_{\beta}(\mu',\B r_2) \rangle_{\B w} \rangle_{\B
v} \,d \mu' \ .
\end{eqnarray*}
\REF{C15} allows to rewrite  \REF{C14} as
\begin{eqnarray} \nn
\Phi(t,\B R) &=& \frac{1}{\bar\tau\sb{ren} }\left[ \int_0^t \exp(\mu
\hat{\cal L}_1) \, d \mu\right]
 \Phi(t,\B R) \\  
&& + \exp(t \hat{\cal L}_1) \Phi(t,\B R) \,,
    \label{eq:C20}
\end{eqnarray}
where
\begin{eqnarray}
  \label{eq:L1}
  \hat{\C L}_1= \hat{\C L} - \frac{\p}{\p t}- {1 \over
  \bar\tau\sb{ren}} \ .
\end{eqnarray}
To derive \REF{C20} we used the following identity
\begin{eqnarray}
\Phi(t-\mu,\B R) = \exp \biggl(- \mu {\partial \over \partial t}
\biggr) \Phi(t,\B R) \;, \label{C1}
\end{eqnarray}
which follows from the Taylor expansion
\begin{eqnarray}
f(t + \tau) = \sum_{m=1}^{\infty} \biggl(\tau {\partial \over \partial
t}
\biggr)^{m} {f(t) \over m!} = \exp \biggl(\tau {\partial
\over \partial t}
\biggr) f(t) \; .
 \label{TC21}
\end{eqnarray}
 In particular,
$$\Phi_0(\B R)=\Phi(t-t,\B R)=\exp\left( - t \frac{\partial}{
 \partial t}\right) \Phi(t,\B R)\ .
$$
Evaluating the integral in \REF{C20} we obtain
\begin{eqnarray}
  \label{eq:TC22}
 \left[\exp(t  \hat{\cal L}_1) -1\right ]\left( \hat{\C L}_1+
     \bar\tau^{-1}\sb{ren}\right) \Phi(t,\B R)=0\ .
\end{eqnarray}
Here we used the commutativity relation
$$ 
\hat{\cal L}_1 \exp(t
\hat{\cal L}_1)= \exp(t \hat{\cal L}_1)\, \hat{\cal L}_1\ .
$$ 
 Thus, finally
\begin{eqnarray}  \nn
&& \hskip -0.5cm {\partial \Phi \over \partial t}=\Big[ B(\B R) +
2 {\bf U}(\B R) \cdot \B {\nabla} + \hat D_{\alpha \beta}(\B R)
\nabla_{\alpha} \nabla_{\beta}
\Big] \Phi(t,\B R)\,,  \\
\label{eq:C22}
\end{eqnarray}
\noindent Note that in the limit $\bar\tau\sb{ren}\to \infty $,
\REF{C22} describes the evolution of $\Phi (t,\B R)$ in the model
of the random velocity field without renewals.
\S{modelC}{Gaussian velocity field with a small yet finite
correlation time}
Here we consider a random Gaussian velocity field with a small $
\bar\tau\sb{ren}$. Using Eq. (\ref{C1}) we rewrite \REF{C14} in
the form
\begin{equation}
\biggl \{ {1 \over \bar\tau\sb{ren}} \int\limits_{0}^{t} \hat
P(\tau,\B R) \exp \biggl(- {\tau \over \bar\tau\sb{ren}} \hat M
\biggr) \, d \tau - 1 \biggr\} \Phi(t,\B R) = 0 \,,
\label{N21}
\end{equation}
where $ \hat M = 1 + \bar\tau\sb{ren} (\partial / \partial t) $
and we neglected the last term in \REF{C14} for small~$\tau$.
Expanding the function $ \hat P(\tau,\B R) $ in Taylor series in
the vicinity of $ \tau = 0 $ we obtain
\begin{eqnarray}
\Big\{\sum_{k=0}^{\infty} \bar\tau^k\sb{ren} \Big[{\partial^k
\hat P(\tau,\B R) \over \partial \tau^k} \Big]_{\tau=0} \hat
M^{-(k+1)} - 1 \Big\} \Phi(t,\B R) = 0 \,, \nonumber\\
\label{ON22}
\end{eqnarray}
where we used that
$$
\int\limits _0^t \tau^k \exp \biggl(- {\tau \over \bar\tau\sb{ren}} \hat
M
\biggr) \, d \tau = k! \, \bar\tau^{k+1}\sb{ren} \hat M^{-(k+1)} \ .
$$
Neglecting the terms $ \sim O(\bar\tau^5\sb{ren}) $ in Eq.
(\ref{ON22}) we obtain
\begin{eqnarray}
\label{ON1} \hat M^{2} {\partial \Phi(t,\B R) \over \partial t} &&
= \bar\tau\sb{ren} \biggl[ \left({\partial^2 \hat P(\tau,\B R)
\over
\partial \tau^2} \right)_{\tau=0}
\\
&& +  \bar\tau^{2}\sb{ren} \left({\partial^4 \hat P(\tau,\B R)
\over \partial \tau^4} \right)_{\tau=0} \biggr] \Phi(t,\B R) \,,
\nonumber
\end{eqnarray}
where we used that the expansion of the operator $ \hat P(\tau,\B
R) $ into Taylor series (for small $ \tau )$ for a random Gaussian
velocity field has only even powers of $ \tau .$ Thus, the
equation for the correlation function $ \Phi(t,\B R) $ is given by
\Onecol
\begin{eqnarray}
\label{N22}
\hat M^{2} \frac{\partial \Phi(t,\B R)}{\partial t}
&=& [B(\B R) + 2 \B U(\B R) {\bf \cdot} \B {\nabla} + \hat
D_{\alpha \beta}(\B R) \nabla_{\alpha} \nabla_{\beta}] \Phi \,,
\end{eqnarray}
where
\begin{eqnarray}
\hat D_{\alpha \beta}(\B R) &=& \frac{\bar\tau\sb{ren}}{2} \lla
\tilde \xi_{\alpha} \tilde \xi_{\beta} G(\B r_1) G(\B r_2)
\rangle_{\B w\B w} \rangle_{\B v} \,,
\label{N1} \\
U_{\alpha}(\B R) &=& - \frac{1}{\bar\tau\sb{ren}} \lla  g(\B r_2)
\xi_{\alpha}^{\ast}(\B r_1) \rangle_{\B w\B w} \rangle_{\B v}
 + \frac{1}{2 \bar\tau\sb{ren}} \lla  g(\B r_1)
g(\B r_2) \tilde \xi_{\alpha}  \rangle_{\B w\B w} \rangle_{\B v}
\;, \label{N2} \\   \label{N3} B(\B R) &=&
\frac{1}{\bar\tau\sb{ren}} \lla g(\B r_1) g(\B r_1) \rangle_{\B
w\B w} \rangle_{\B v} \;,
\end{eqnarray}
Here for the homogeneous turbulent velocity field \cite{EKRS99}:
\[
\tilde {\B \xi} = \B {\xi}'(\B r_2) - \B {\xi}'(\B r_1)\,, \qquad
\B {\nabla}= \partial / \partial \B R\,, \qquad
 G =\bar G + g ,\qquad
\langle \langle g \rangle_{\B w\B w} \rangle_{\B v} = 0 \,, \qquad
\bar G = \langle \langle G \rangle_{\B w\B w} \rangle_{\B v} = 1\ .
\]
Using the expansion of $ \B {\xi}(\bar\tau\sb{ren},\B r) $ and $
g[\bar\tau\sb{ren},\B {\xi}(\B r)]$ into Taylor series of a small
time $ \bar\tau\sb{ren} $ after the lengthly algebra we obtain
\begin{eqnarray}
\hat D_{\alpha \beta}(\B R) &=& 2 D \delta_{\alpha \beta} + 2
\bar\tau\sb{ren} [\tilde f_{\alpha \beta}(\B R) + {\rm St}^{2}
Q_{\alpha \beta}(\B R)] \;,
\label{N5} \\  
Q_{\alpha \beta}(\B R) &=& 3 [(\nabla_{\nu} f_{\mu \beta})
(\nabla_{\mu} f_{\alpha \nu}) - \tilde f_{\mu \nu} \nabla_{\nu}
\nabla_{\mu} f_{\alpha \beta} ] + 24 A_{\alpha} A_{\beta} + 12
(A_{\mu} \nabla_{\mu} f_{\alpha \beta} - \tilde f_{\alpha \beta}
\nabla_{\mu} A_{\mu}) - 20 \tilde f_{\alpha \mu} \nabla_{\mu}
A_{\beta} \;,
\label{N6} \\
U_{\alpha}(\B R) &=& - 2 \bar\tau\sb{ren} \{ A_{\alpha} - {\rm
St}^{2} [(\nabla_{\nu} A_{\mu}) (\nabla_{\mu} f_{\alpha \nu}) + 10
A_{\mu} \nabla_{\mu} A_{\alpha} + 12 A_{\alpha} \nabla_{\mu}
A_{\mu} ] \} \;,
\label{N7} \\
B(\B R) &=& - 2 \bar\tau\sb{ren} \{\nabla_{\mu} A_{\mu} + {\rm
St}^{2} [(\nabla_{\nu} A_{\mu}) (\nabla_{\mu} A_{\nu}) - 6
(\nabla_{\mu} A_{\mu})^{2} ] \} \;, \label{N8}  \br A_{\alpha} &=&
\nabla_{\beta} f_{\alpha \beta} \,, \quad \tilde f_{\alpha \beta}
= f_{\alpha \beta}(0) - f_{\alpha \beta}(\B R)\,, \quad f_{\alpha
\beta}(\B R) = \langle v_{\alpha}(\B r_1) v_{\beta}(\B r_2)
\rangle_{\B v}\,,
\end{eqnarray}
\Twocol \noindent 
and St$ = \bar\tau\sb{ren} / \tau_{\eta} $ is the Strouhal number.
In these calculations we neglected small terms $ \sim O({\rm St}^2
R^3 \nabla^3) .$ Our analysis showed that the neglected small
terms do not affect the growth rate of the clustering instability.
In Eqs. (\ref{N6})-(\ref{N8}) we assumed that the correlation
function $ f_{\alpha \beta} $ for homogeneous, isotropic and
compressible velocity field is given by
\begin{equation}
f_{\alpha \beta}(\B R) = \frac{u_{\eta}^{2}}{3} \Big[ (F + F_c)
\delta_{\alpha \beta} + \frac{R F'}{2} P_{\alpha \beta} + R F'_c
R_{\alpha \beta} \Big]\,,
\label{N9}
\end{equation}
(see \cite{EKR95}), and in scales $ 0 < R \ll 1 $ incompressible $
F(R) $ and compressible $ F_c(R) $ components of the random
velocity field are given by
\[
F(R) = (1 - R^{2}) / (1 + \sigma_v)\,,\quad
F_c(R) = \sigma_v F(R)\,,
\]
in scales $ R \geq 1 $ the functions $ F = F_c = 0 .$ Here $ R $ is
measured in the units of $ \eta ,$ $ \quad P_{\alpha \beta}(R) =
\delta_{\alpha \beta} - R_{\alpha \beta} ,$ $ \quad R_{\alpha \beta} =
R_{\alpha} R_{\beta} / R^{2} $ and $ F' = dF / dR .$ Turbulent
diffusion tensor $ D_{\alpha \beta}(R) $ is determined by the field of
Lagrangian trajectories $ \B {\xi} $ [see Eq. (\ref{N1})]. Due to a
finite correlation time of a random velocity the field of Lagrangian
trajectories $ \B {\xi} $ is compressible even if the velocity field
is incompressible $ (\sigma_v = 0 ).$ Indeed, for $ \sigma_v = 0 $ we
obtain
$$ 
 \langle \langle (\B {\nabla} \cdot \B {\xi})^{2} \rangle_{\B
w} \rangle_{\B v} = \frac{20}{3} {\rm St}^{4}\ .
$$ 
  Using Eqs. (\ref{N5})-(\ref{N9}) we calculate the functions $ \hat
  D_{\alpha \beta}(\B R) ,$ $ \quad U_{\alpha}(\B R) $ and $ B(\B R)
  :$
\begin{eqnarray}   \nn
\hat D_{\alpha \beta}(\B R)&=&[2 D + R^{2}(a_{3} + {\rm St}^{2}
b_{6})] \delta_{\alpha \beta}
\\
&& + R^{2} (a_{4} + {\rm St}^{2} b_{4}) R_{\alpha \beta},
\\
U_{\alpha}(\B R) &=& - R_{\alpha} (a_{5} + {\rm St}^{2} b_{5}) \,,
\\
B &=& a_{6} + {\rm St}^{2} b_{2} \;, \label{N12}
\end{eqnarray}
where $ b_{2} = - \frac{51}{47} b_{5}$ and
\begin{eqnarray*}
a_{5} &=& - \frac{20 \sigma_v}{3(1 + \sigma_v)} = -
\frac{a_{2}}{3} \;, \quad a_{3} = \frac{2 \sigma_v + 4}{3(1 +
\sigma_v)} \;,
\\
a_{4} &=& {4 \sigma_v - 2 \over 3(1 + \sigma_v)} \;, \quad b_{5} =
- {2350 \over 27} \biggl({\sigma_v \over 1 + \sigma_v} \biggr)^{2}
\;,
\\
b_{6} &=& \frac{12 + 872 \sigma_v + 433 \sigma_v^{2} }{27 (1 +
\sigma_v)^{2}}  \;,
\\
b_{4} &=& \frac{2 (12 - 203 \sigma_v
+ 1033 \sigma_v^{2}) }{27 (1 + \sigma_v)^{2}} \; .
\end{eqnarray*}
 We will show here that the combined effect of particles inertia $
 (\sigma_v \not= 0) $ and finite correlation time $ ({\rm St} \not =
 0) $ results in the excitation of the clustering instability whereby
 under certain conditions there is a self-excitation of the second
 moment of a number density of inertial particles. This instability
 causes formation of small-scale inhomogeneities in spatial
 distribution of inertial particles.

The equation for the second-order correlation function for the number
density of inertial particles reads
\begin{equation}
\hat M^{2} \frac{\partial \Phi(t,R)}{\partial t} = \frac{
\Phi''}{m(R)} + \tilde \lambda(R) \Phi' + B \Phi \; \label{N14}
\end{equation}
[see Eqs.~(\ref{N12})], where the time $t$ is measured in units of
$ t_{\eta} ,$ and
\begin{eqnarray*}
\Phi' &\!\! =\!\!& \frac{\partial \Phi}{\partial R}\,,\qquad
\Phi'' = \frac{\partial^{2} \Phi}{\partial R^{2}}\,,\qquad
\frac{1}{m} = \frac{ 2 (1 + X^2)}{\text{Sc}}\,,
\\
\tilde \lambda &\!\! =\!\!&  \frac{2 [2 + X^{2}(1 + 2 C)]}{R \,
 {\rm Sc}}\,,  \qquad \quad
C =\frac{a_{1} + {\rm St}^{2} b_{1}}{4 \beta}\,, \\
\beta  &\!\! =\!\!&  \frac{a_{2} + {\rm St}^{2} b_{3}}{2}\,, \
X(R)=\sqrt{{\rm Sc} \beta} \, R , \  R = \vert \B
r_2 - \B r_1 \vert\,,\ \\
a_{1} &=& {2(19 \sigma_v + 3) \over 3(1 + \sigma_v)} \;,   \qquad
\qquad a_{2} = {2(3 \sigma_v + 1) \over 3(1 + \sigma_v)} \;,
\\
b_{1} &=& - {1 \over 27 (1 + \sigma_v)^{2}} (12 - 1278 \sigma_v -
3067 \sigma_v^{2}) \;,
\\
b_{3} &=& {1 \over 27 (1 + \sigma_v)^{2}} (36 + 466 \sigma_v +
2499 \sigma_v^{2}) \ .
\end{eqnarray*}
In order to obtain a solution of Eq. (\ref{N14}) we use a
separation of variables, i.e., we seek for a solution in the
following form:
\[\Phi(t,R) = \hat \Phi(R) \exp(\gamma_2 t)\,,\]
whereby $ \gamma_2 $ is a free parameter which is determined using
the boundary conditions
$$ 
 \hat \Phi(R=0) = 1\,, \quad \hat \Phi(R
\to \infty) = 0\ .
$$ 
 Here $ \gamma_2 $ is measured in units of $ 1/t_{\eta}$. Since the
 function $ \Phi(t,R) $ is the two-point correlation function, it has
 a global maximum at $ R=0 $ and therefore it satisfies the
 conditions:
\begin{eqnarray*}
\hat \Phi' (R=0) &=& 0\,, \quad \hat \Phi'' (R=0) < 0 \,,
\\
\hat \Phi (R=0) & >&  | \hat \Phi (R>0)|\ .
\end{eqnarray*}
Then Eq. (\ref{N14}) yields
\begin{eqnarray}
\Gamma \hat \Phi(R) = {1 \over m(R)} \hat \Phi'' +
\tilde \lambda(R) \hat \Phi' + B \hat \Phi \,,
\label{NN14}
\end{eqnarray}
where $ \Gamma = \gamma_2 (1 + \bar\tau\sb{ren} \gamma_2)^{2}$ .
Equation (\ref{NN14}) has an exact solution for $ 0 \leq R < 1 :$
\begin{eqnarray}
\hat \Phi(X) &=& {S(X)}{X (1 + X^2)^{\mu/2}} \;, \label{N15} \br
S(X) &=& \text{Re} \{ A_1 P_\zeta^\mu(iX) + A_2 Q_\zeta^\mu(iX) \}
\,,
 \end{eqnarray}
and $ P_\zeta^\mu(Z) $ and $ Q_\zeta^\mu(Z) $ are the Legendre
functions with imaginary argument
$$ 
 Z = i X\,, \ \mu = C -
\frac{3}{2}\,, \ \zeta = - \hf \pm \sqrt{C^{2} - \kappa}\,, \
 \kappa = \frac{B - \Gamma}{2 \beta}.
$$

Solution of Eq. (\ref{N14}) can be analyzed using asymptotics of
the exact solution (\ref{N15}). This asymptotic analysis is based
on the separation of scales (see, e.g.,   \cite{ZMR88,EKR95}).  In
particular, the solution of Eq. (\ref{N14}) has different regions
where the form of the functions $ m(R) $ and $ \tilde \lambda(R) $
are different. The functions $ \hat \Phi(R) $ and $ \hat \Phi'(R)
$ in these different regions are matched at their boundaries in
order to obtain continuous solution for the correlation function.
Note that the most important part of the solution is localized in
small scales (i.e., $ R \ll 1 ) .$ Using the asymptotic analysis
of the exact solution for $ X \gg 1 $ allowed us to obtain the
necessary conditions of a small-scale instability of the second
moment of a number density of inertial particles.  The results
obtained by this asymptotic analysis are presented below.

The solution (\ref{N15}) has the following asymptotics: for $ X
\ll 1 $ (i.e., in the scales $ 0 \leq R \ll 1 / \sqrt{\rm Sc}) $
the solution for the second moment $ \hat \Phi $ is given by
\begin{eqnarray}
\hat \Phi(X) = \{1 - (\kappa / 6) [X^{2} + O(X^{4})] \} \; .
\label{N16}
\end{eqnarray}
For $ X \gg 1 $ (i.e., in the scales $ 1 / \sqrt{\rm Sc} \ll R <
1) $ the function $ \hat \Phi $ is given by
\begin{equation}\label{blya}
 \hat \Phi(X) =
\text{Re} \{ A X^{- C \pm \sqrt{C^{2}-\kappa}} \} .
\end{equation} 
 When $ C^{2}-\kappa < 0 $ the second-order correlation function for a
 number density of inertial particles $ \hat \Phi $ is given by
$$ 
 \hat
 \Phi(R) = A_{3} R^{-C} \cos(\nu_{I} \ln R + \varphi)\,, \quad
 \nu_{I} = \sqrt{\kappa - C^{2}}\,,
$$ 
where $ C > 0 $ and $ \varphi $ is
 the argument of the complex constant $ A .$ For $ R \geq 1 $ the
 second-order correlation function for the number density of inertial
 particles is given by
\begin{eqnarray}
\hat \Phi(R) = (A_{4} /  R) \exp(- R \, \sqrt{3 \Gamma / 2}) \;,
\label{N19}
\end{eqnarray}
where $ \Gamma > 0$. Since the total number of particles in a
closed volume is conserved, \ie, particles can only be
redistributed in the volume,
$$ 
 \int_{0}^{\infty} R^{2} \hat
\Phi(R) \,dR = \hat \Phi(k=0) = 0\ .
$$ 
 The latter yields $ \varphi = - \pi / 2 $ for $ \ln {\rm Sc} \gg 1 $
 and $ \Gamma \ll 1 .$ When $ C^{2}-\kappa > 0 $, the solution
 (\ref{blya}) cannot be matched with solutions (\ref{N16}) and
 (\ref{N19}).  Thus, the condition $ C^{2}-\kappa < 0 $ is the
 necessary condition for the existence of the solution for the
 correlation function. The condition $ C > 0 $ provides the existence
 of the global maximum of the correlation function at $ R=0$.

Matching functions $ \hat \Phi $ and $ \hat \Phi' $ at the
boundaries of the above-mentioned regions yields coefficients $
A_{k} $ and $ \Gamma .$ In particular, the eigenvalue $ \Gamma $
is given by Eq. (\ref{N20}).



\end{document}